\documentclass[lettersize,journal]{IEEEtran}
\usepackage{amsmath,amsfonts}
\usepackage{algorithmic}
\usepackage{array}
\usepackage{textcomp}
\usepackage{stfloats}
\usepackage{url}
\usepackage{verbatim}
\usepackage{graphicx}
\usepackage{booktabs} 
\usepackage{subcaption}

\usepackage{bm}
\usepackage{multirow}
\usepackage{makecell}
\usepackage{pifont}
\usepackage[pagebackref,breaklinks,colorlinks]{hyperref}
\usepackage{balance}

\newcommand{\ieno}{\textit{i}.\textit{e}.} 
\newcommand{\egno}{\textit{e}.\textit{g}.} 

\def\BibTeX{{\rm B\kern-.05em{\sc i\kern-.025em b}\kern-.08em
    T\kern-.1667em\lower.7ex\hbox{E}\kern-.125emX}}

\begin{document}

\title{Learning Cross-Scale Weighted Prediction for Efficient Neural Video Compression}
\author{Zongyu Guo\IEEEauthorrefmark{1}\thanks{\IEEEauthorblockA{\IEEEauthorrefmark{1} Equal Contributions}}, Runsen Feng\IEEEauthorrefmark{1}, Zhizheng Zhang, Xin Jin and Zhibo Chen\IEEEauthorrefmark{2},~\IEEEmembership{Senior Member,~IEEE,}\thanks{
\IEEEauthorblockA{\IEEEauthorrefmark{2} Corresponding Author: Zhibo Chen (chenzhibo@ustc.edu.cn).} 

\IEEEauthorblockA{Zongyu Guo (email: guozy@mail.ustc.edu.cn), Runsen Feng, Xin Jin and Zhibo Chen are with the Department of Electronic Engineer and Information Science, University of Science and Technology of China, Hefei, Anhui, China. Zhizheng Zhang is with Microsoft Research Asia, Beijing, China. }}}

\markboth{Preprint}%
{Shell \MakeLowercase{\textit{et al.}}: A Sample Article Using IEEEtran.cls for IEEE Journals}

\maketitle

\begin{abstract}

Neural video codecs have demonstrated great potential in video transmission and storage applications. Existing neural hybrid video coding approaches rely on optical flow or Gaussian-scale flow for prediction, which cannot support fine-grained adaptation to diverse motion content. Towards more content-adaptive prediction, we propose a novel cross-scale prediction module that achieves more effective motion compensation. Specifically, on the one hand, we produce a reference feature pyramid as prediction sources and then transmit cross-scale flows that leverage the feature scale to control the precision of prediction. On the other hand, for the first time, a weighted prediction mechanism is introduced even if only a single reference frame is available, which can help synthesize a fine prediction result by transmitting cross-scale weight maps. 
In addition to the cross-scale prediction module, we further propose a multi-stage quantization strategy, which improves the rate-distortion performance with no extra computational penalty during inference. We show the encouraging performance of our efficient neural video codec (ENVC) on several benchmark datasets. In particular, the proposed ENVC can compete with the latest coding standard H.266/VVC in terms of sRGB PSNR on UVG dataset for the low-latency mode. We also analyze in detail the effectiveness of the cross-scale prediction module in handling various video content, and provide a comprehensive ablation study to analyze those important components. Test code is available at \url{https://github.com/USTC-IMCL/ENVC}.

\end{abstract}


\section{Introduction}
\label{sec:intro}

\IEEEPARstart{T}{he}  ML-based methods have shown promise in reshaping the field of data compression. The success in neural image compression \cite{balle2018variational,cheng2020learned,mentzer2020high,guo2021causal} clearly demonstrates the coding efficiency of ML-based methods. But video compression is a more challenging task, due to the complicated inter frame redundancies. The evolutionary history of traditional video coding standards \cite{wiegand2003overview,sullivan2012overview,bross2021overview} proves the significance of inter frame prediction  \cite{zhang2013background,zhang2018improved}. A good prediction module should be aware of the content variety of video frames and be adaptive to diverse motions. 

Many neural video codecs have been designed in the past few years. Some are proposed for the low delay setting, reducing the temporal redundancies by unidirectional prediction \cite{lu2019dvc,chen2019learning,habibian2019video,agustsson2020scale,lin2020m,liu2020conditional,hu2021fvc,liu2021neural,rippel2021elf}. Some can be used for the random access setting, making use of bidirectional references to predict the target frame \cite{wu2018video,djelouah2019neural,yang2020learning,Pourreza_2021_ICCV}. In this paper, we focus on the simplest low-latency mode, \ieno, always predicting the target frame with the previous single reconstructed frame. This is the most basic scenario of inter frame prediction, which reveals the ability of a video codec in exploiting the temporal redundancies within two consecutive frames. Actually, a successful neural video codec designed for this scenario (single-reference prediction) can easily be extended to other prediction scenarios \cite{lin2020m,Pourreza_2021_ICCV}.

DVC \cite{lu2019dvc} is a typical method designed for this scenario. DVC follows the classical hybrid coding framework \cite{habibi1974hybrid}, compressing the 2-D motion vectors as optical flow and then compressing the pixel residuals by end-to-end optimization. The optical flow describes the motion displacement of every pixel, utilized to generate the warping result (the predicted frame). 
However, prediction with 2-D optical flow fails frequently when there is no reference for frame prediction, such as disocclusion. To this end, the later work SSF \cite{agustsson2020scale} applies multi-level Gaussian smoothing on the reference frame and performs trilinear warping in Gaussian-scale space \cite{lindeberg2013scale}. The scale dimension in SSF allows the model to blur the source content before prediction, reducing the errors caused by the direct warping of optical flow. 
But SSF can only help to smooth the predicted result with stationary blurring patterns, not fully adaptive to diverse video content. Recently, FVC \cite{hu2021fvc} suggests learning reference representations for feature-level prediction, which also reduces prediction artifacts of pixel-level warping. However, regarding the learned reference sources, there is no concept of scale to explicitly control the prediction precision. Such feature-level prediction thereby cannot adapt to video content and fails to handle various motion cases.

\begin{figure}[t]
 \centering
 \includegraphics[scale=0.210, clip, trim=9.5cm 1.5cm 9.6cm 34.5cm]{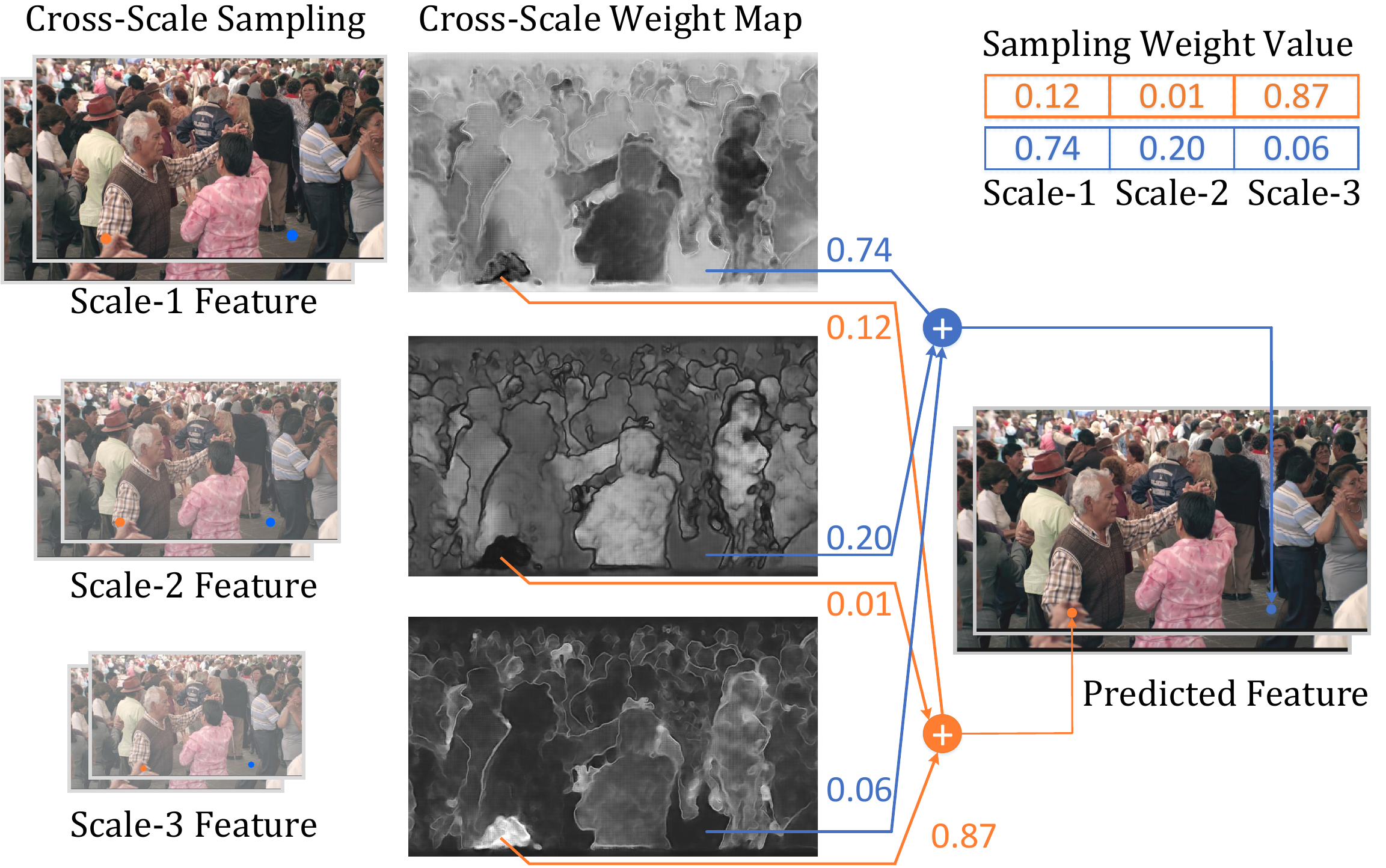}
 \caption{The proposed cross-scale prediction for better motion compensation. Our model generates prediction sources at three scales. Cross-scale flows are transmitted to determine the sampling locations across scales. Corresponding weight maps describe the importance of sampling results.}
\label{figure1}
\end{figure}

In this paper, towards fine-grained content-adaptive prediction, we propose a novel cross-scale prediction module, achieving much more effective motion compensation in the scenario of single-reference prediction. 
While processing features at multiple scales is one of the most successful techniques for computer vision \cite{lin2017feature,zhu2020deformable,Liu_2021_ICCV}, we show that in the task of neural video compression, we can produce a reference feature pyramid  for flexible prediction of diverse motions. 
The high-resolution reference features are well-suited for prediction in the case of simple motion, \egno, translational motion, since the high-resolution features preserve most of the context and deliver a high-precision prediction. Meanwhile, we can generate the low-resolution reference features with strong semantic information, to further reduce the prediction error in many other motion cases (explained in detail in Section \ref{section_3_2}). After producing the reference feature pyramid, our model will transmit cross-scale flows as a set of displacement vectors that work with the multi-scale prediction sources. Our neural video codec can cover multiple prediction precisions to meet the requirements of diverse motion cases.

While producing a reference feature pyramid provides more options on prediction sources, it is still hard to support fine-grained adaptation to diverse motion content. Motivated by previous works that apply weight maps for prediction with multiple reference frames \cite{djelouah2019neural,ladune2021conditional}, we introduce the mechanism of \textit{weighted prediction} to compute a content-adaptive prediction result even in the scenario of single-reference prediction. 
Specifically, we transmit cross-scale weight maps as a part of motion information to describe the importance of each sampling result. We highlight that the reference features learned at multiple scales are complementary to each other in various motion cases. 
As shown in Figure \ref{figure1}, the region with simple motion will pay more attention to the high-resolution prediction result (assigning larger weight value to scale-1, the blue point). The region with complicated motion (the orange point) will be predicted with larger weight value assigned to the low-resolution scale (scale-3 in Figure \ref{figure1}). Therefore, our proposed cross-scale prediction module can adapt to diverse motion content and synthesize a fine prediction result.

In addition, we also propose a multi-stage quantization strategy that only affects the training process of our neural video codec but delivers higher coding efficiency during testing. Such multi-stage quantization strategy helps to close the train-test mismatch, which is an underlying issue of additive uniform noise-based quantization \cite{balle2016end}. We term this strategy as the video-version soft-then-hard, as it is improved from the image-version soft-then-hard \cite{guo2021soft}.

In short, our contributions are summarized as follows:

\begin{itemize}
\item We introduce a novel inter-frame prediction approach for video compression, \ieno, cross-scale weighted prediction. By combining multi-scale reference sources and multi-head warping, this weighted prediction mechanism can adapt to diverse motion cases, making it a flexible and effective solution for video compression with a single reference frame.
\item We propose a multi-stage quantization strategy, improving the RD performance with no extra computational penalty during inference.
\item We carefully build our neural video codec and achieve the encouraging rate-distortion performance.
\end{itemize}

While most previous neural video codecs can barely compete with H.265/HEVC, our proposed efficient neural video codec (ENVC) can compete with H.266/VVC \cite{bross2021overview} in terms of sRGB PSNR under the same prediction configuration (single reference frame, intra period is 12) on UVG dataset \cite{mercat2020uvg}. When optimized for the MS-SSIM metric \cite{wang2004image}, ENVC outperforms VVC obviously. We also show our encouraging performance on other common benchmark datasets, including HEVC test sequences \cite{sullivan2012overview} and MCL-JCV \cite{wang2016mcl}. 
We provide a comprehensive ablation study to analyze the effectiveness of our proposed cross-scale prediction module and other important components that contribute to the state-of-the-art performance.

\section{Related Work}

\subsection{Lossy Image Compression}

Despite a short history, neural image codecs have outperformed the latest image compression standard H.266/VVC intra regarding sRGB PSNR \cite{guo2021causal} and MS-SSIM \cite{cheng2020learned}. Currently, most prevailing neural image codecs follow the VAE framework \cite{balle2016end,balle2018variational}. A series of works are built upon this framework, improving from the aspects of entropy estimation \cite{minnen2018joint,Lee2019Context,cheng2020learned,guo2021causal,chen2021end}, quantization \cite{balle2016end,agustsson2017soft,yang2020improving,agustsson2020universally,guo2021soft}, variable rate \cite{choi2019variable,cui2021asymmetric} and perceptual quality \cite{agustsson2019generative,mentzer2020high}. Among them, we note that the autoregressive (AR) context model \cite{minnen2018joint,Lee2019Context} can achieve obvious rate savings but bring much more decoding complexity. In this paper, we build two versions of video codecs, with or without the AR model. The former achieves the best rate-distortion (RD) performance while the latter is more applicable in practice.

\subsection{Video Compression}

It takes decades to develop a new generation of video coding standard. The classical standards H.264/AVC \cite{wiegand2003overview} and H.265/HEVC \cite{sullivan2012overview} are in broad use, and the recent standardized H.266/VVC \cite{bross2021overview} will soon be applied in industry. All of these standards follow the hybrid coding framework, developed with handcrafted modes. Technically, in a video codec, the intra frame coding is equivalent to image coding. And the challenge for video compression mainly exists in efficient compression of inter frames \cite{zhang2013background,zhang2018improved}.

Neural video compression approaches are on the way catching up with traditional standards. Existing works in this field can be classified into two categories, designed for low delay setting \cite{lu2019dvc,chen2019learning,habibian2019video,agustsson2020scale,lin2020m,liu2020conditional,hu2021fvc,liu2021neural,rippel2021elf} and random access setting \cite{wu2018video,djelouah2019neural,yang2020learning,Pourreza_2021_ICCV,yilmaz2021end}. The low delay setting is suitable for applications such as live streaming, which only uses the past frame(s) to predict the current frame. For the random access setting, the reference frame can be from future, well-suited for applications such as playback. In addition, there are works designing a unified model for both settings \cite{ladune2021conditional,feng2021versatile}. 

The compression model applied for the low delay setting is termed as a P-frame model. Many techniques have been proposed to improve the coding efficiency of P-frame model, such as 3D autoencoders \cite{habibian2019video}, conditional entropy coding \cite{habibian2019video,liu2020conditional,li2021deep}, online updating \cite{lu2020content} and effective prediction in hybrid coding. For these hybrid coding solutions, they further involve two prediction modes, \ieno, prediction with single or multiple reference frames. The aforementioned DVC \cite{lu2019dvc}, SSF \cite{agustsson2020scale} and FVC \cite{hu2021fvc} are typical works improving single-reference prediction. Some multi-frame fusion modules are also designed for unidirectional prediction with multiple reference frames \cite{lin2020m,hu2021fvc,rippel2021elf}. Obviously, fusing more reference frames benefits the RD performance, but also brings a significant increase of memory cost.

A successful P-frame model designed for single-reference prediction can easily be extended to unidirectional multiple-reference prediction or bidirectional prediction. Therefore, prediction only with the previous frame is the most basic case for video compression, as the main concern of our paper.

\subsection{Concurrent Work}

There are several works \cite{liu2021neural,sheng2022temporal,hu2022coarse,li2022hybrid} that are almost concurrent with ours. Here, we provide detailed comparisons with these papers, highlighting our unique contributions and superiorities.

The work of \cite{liu2021neural}, which was publicly released a few of months earlier than our work, adopts multi-scale motion compensation for inter-frame prediction. The pyramid flow decoder in \cite{liu2021neural} is trained to pursue accurate prediction in all feature levels, as illustrated in Equation 3 of their paper. In contrast, we tend to believe that the reference features in different levels should play different roles for inter-frame prediction, and we should not pursue reconstructing features perfectly in all levels at the decoder sider. Our proposed cross-scale weighted prediction mechanism helps to integrate information from all feature levels and learn effective reference features freely. Two other concurrent works, \cite{sheng2022temporal} and \cite{hu2022coarse}, also employ the concept of multi-scale motion compensation. Specifically, similar to \cite{liu2021neural}, \cite{sheng2022temporal} decodes out the multi-scale flow maps at different feature levels and then send the multi-scale temporal context into a feature pyramid network (FPN) inside a contextual encoder that replaces the residual encoder in our paper. On the other hand, \cite{hu2022coarse} explicitly transmits two groups of motion information for coarse and fine motion compensation, where redundancies may exist within these two motion information groups. 

An important difference between our work and these three papers is our design of the weighted prediction mechanism. In our approach, both the flows and the weight maps are transmitted together as motion information, which is a concise but more flexible design with potentially fewer network parameters and computations. Another paper \cite{li2022hybrid}, published a couple of months after \cite{hu2022coarse}, improves the entropy model and adopts a multi-granularity quantization strategy. This paper demonstrates impressive rate-distortion performance when the GoP size is set to 32. Since our paper primarily focuses on the motion compensation part, the contributions of \cite{li2022hybrid} are almost orthogonal to ours.

\section{Method}

\subsection{Overview}
\label{section_3_1}

\begin{figure}[t]
 \centering
 \includegraphics[scale=0.63, clip, trim=5.4cm 6.6cm 5cm 1.6cm]{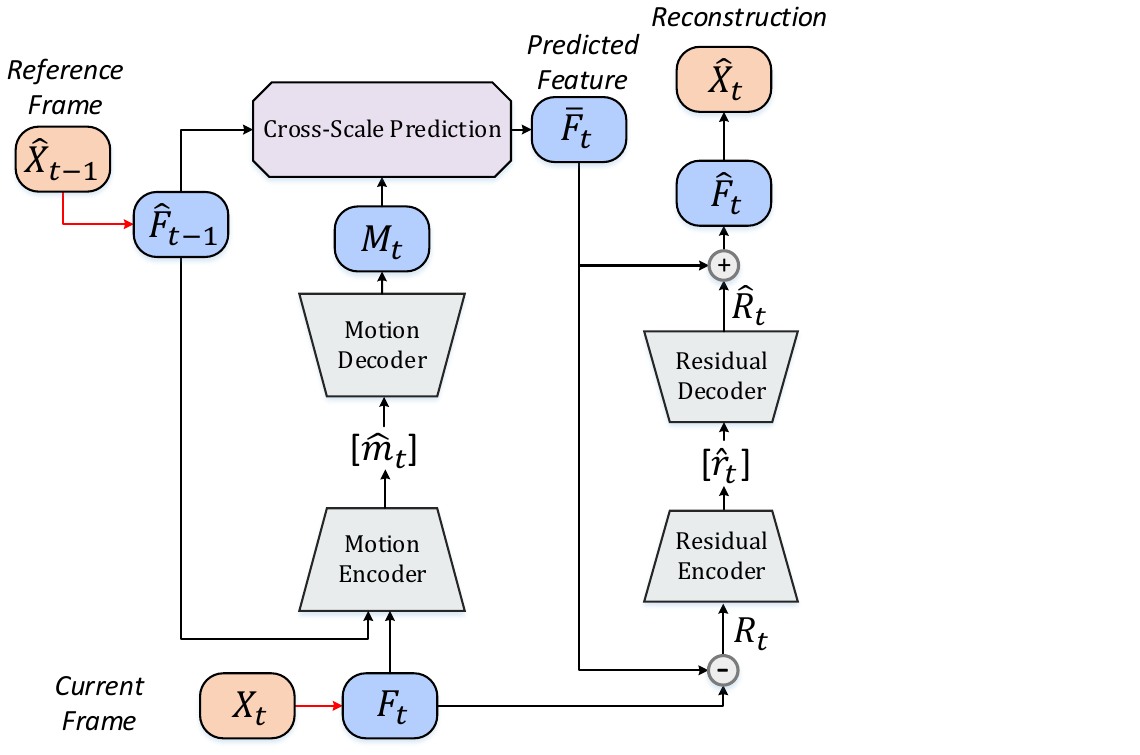}
 \caption{An overview of our P-frame compression model, applied for the low latency mode with single-reference prediction. The proposed cross-scale prediction module (marked in purple) plays an important role for content-adaptive prediction. All blue features are at the scale of $1/2$ original resolution.}
\label{figure2}
\end{figure}

An overview of our P-frame compression model is shown in Figure \ref{figure2}, which follows a typical hybrid coding framework \cite{habibi1974hybrid}. Our P-frame model mainly contains a motion compression network and a residual compression network, both are in the style of autoencoders \cite{balle2016end}. Here, we introduce them respectively by describing the scale (spatial resolution) of some important inputs and outputs inside the framework. The detailed network structures for down/up-sampling and encoding/decoding will be described later in Section \ref{section_3_3}.
\\[3pt]
\textbf{Motion Compression.} 
Given the reference frame $\hat{X}_{t-1}$, which is the lossy reconstruction of the previous frame, it will first be transformed into reference feature $\hat{F}_{t-1}$ by a feature extractor (denoted by the red arrow in Figure \ref{figure2}). The current frame $X_t$ will also be transformed into feature $F_t$, using the same feature extractor. Assume the spatial resolution of $X_t$ is $H\times W$, then both $\hat{F}_{t-1}$ and $F_t$ are features at scale $\frac{H}{2}\times \frac{W}{2}$. $\hat{F}_{t-1}$ and $F_t$ will be concatenated and sent into a motion encoder, generating the quantized variables $\hat{m}_t$ at scale $\frac{H}{16}\times \frac{W}{16}$. On the decoder side, the motion decoder will upsample $\hat{m}_t$ for three times, outputting $M_t$ with scale $\frac{H}{2}\times \frac{W}{2}$. Here, $M_t$ refers to the received motion information, which will participate in cross-scale prediction with the original reference feature $\hat{F}_{t-1}$. The mechanism of this prediction module will be introduced later in Section \ref{section_3_2}.
\\[3pt]
\textbf{Residual Compression.}
The result of cross-scale prediction is the predicted feature $\bar{F}_t$ at scale $\frac{H}{2}\times \frac{W}{2}$. We then compute the feature-level residual \cite{feng2020learned} as
\begin{equation}
R_t = F_t - \bar{F}_t.
\end{equation}
The feature residual $R_t$ will be downsampled in a residual encoder and is then quantized into latent variables $\hat{r}_t$. A residual decoder will decode out $\hat{R}_t$ at the scale of $\frac{H}{2}\times \frac{W}{2}$. Now we can reconstruct the target feature $\hat{F}_t$:
\begin{equation}
\hat{F}_t = \bar{F}_t + \hat{R}_t.
\end{equation}
Finally, a simple upsampling module will transform $\hat{F}_t$ into $\hat{X_t}$, as the final reconstructed frame.

\subsection{Cross-Scale Prediction}
\label{section_3_2}

Here, we introduce the novel cross-scale prediction module in detail. In Figure \ref{figure2}, we can see the inputs of this module include the original reference feature $\hat{F}_{t-1}$ and the received motion information $M_t$. The output is the predicted feature $\bar{F}_t$. All of them are at the scale of $\frac{H}{2}\times \frac{W}{2}$. However, inside the prediction module, features are processed at three scales.

\begin{figure}[t]
 \centering
\begin{subfigure}{\columnwidth}
 \includegraphics[scale=0.35, clip, trim=3.3cm 25.4cm 1.3cm 4.9cm]{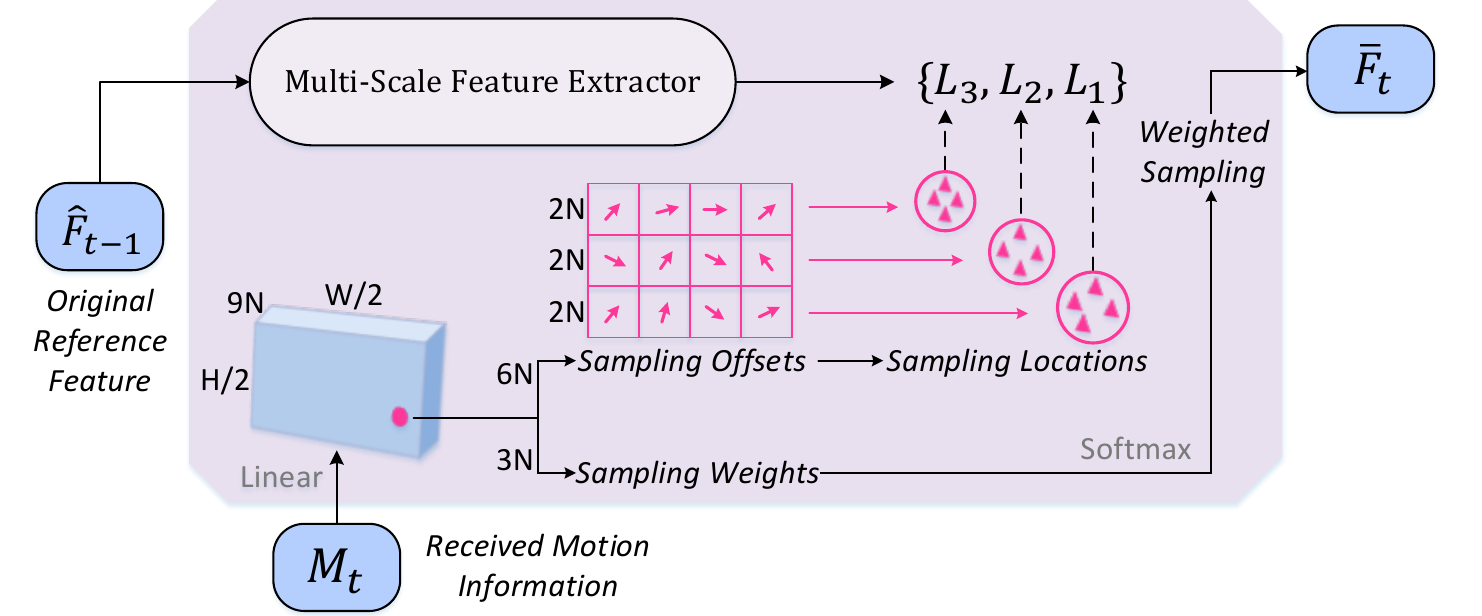}
\caption{}
\label{figure3a}
\end{subfigure}
\begin{subfigure}{\columnwidth}
 \includegraphics[scale=0.37, clip, trim=3.3cm 11.9cm 1.3cm 20.3cm]{./figures/figure3.pdf}
\caption{}
\label{figure3b}
\end{subfigure}
 \caption{(a) Our proposed cross-scale prediction module. (b) The weighted prediction mechanism that can aggregate cross-scale sampling results adaptively.}
\label{figure3}
\end{figure}

As shown in Figure \ref{figure3a}, the original reference feature $\hat{F}_{t-1}$ will be sent into a multi-scale feature extractor to generate a reference feature pyramid $\{L_3,L_2,L_1\}$, The scales of $L_3,L_2,L_1$ are $\frac{H}{8}\times \frac{W}{8}$, $\frac{H}{4}\times \frac{W}{4}$ and $\frac{H}{2}\times \frac{W}{2}$.
The received motion information $M_t$ will pass through a linear projection layer to generate sampling information. Every sample is described with a 3-channel vector, including the horizontal and vertical displacement, and a corresponding weight value. 
At each scale, the reference feature is sampled for $N$ times. Since we have three scales in total, there are $(3N)$ sampling results for every target spatial location. And the sampling information should be a tensor with size $(9N) \times \frac{H}{2}\times \frac{W}{2}$, including $(6N) \times \frac{H}{2}\times \frac{W}{2}$ as the sampling offsets (named \textit{cross-scale flows}) and $(3N) \times \frac{H}{2}\times \frac{W}{2}$ as the sampling weights. All the $(3N)$ sampling results are aggregated according to the softmax-computed sampling weights (named \textit{cross-scale weight maps}). Figure \ref{figure3b} illustrates the aggregation process when $N$ is 4, as our final setting.

The above introduction is a simplified version of our model. Inspired by \cite{zhu2020deformable}, we further apply multi-head warping, where we first project the reference feature pyramid into four channel groups. In each channel group, our model repeats the same operation of cross-scale prediction. As a result, we will obtain multiple groups of predicted features, which are then integrated by another linear projection layer. The visualization in Figure \ref{figure3} can be regarded as the case when head number is 1. We set head number as 4 in our final model, which is found to accelerate the convergence during training. That is to say, the actual sampling information is described as a tensor with shape $(9N \times 4) \times \frac{H}{2}\times \frac{W}{2}$.
\\[4pt]
\textbf{Cross-scale weighted prediction.} The proposed prediction module is \textit{cross-scale} due to two reasons. First, the prediction sources come from a multi-scale feature pyramid but the prediction result is at the single scale of $\frac{H}{2}\times \frac{W}{2}$. It means the cross-scale prediction module itself can aggregate information from multi-scale reference features, without the help of a down-to-top FPN structure \cite{lin2017feature}. Second, cross-scale weight maps are also transmitted, which help to generate the prediction result at the single scale from multiple reference scales. Therefore, this cross-scale weighted prediction module is beyond the standard feature pyramid. We also should note that although the \textit{weighted} prediction mechanism is common to merge the information from multiple reference frames \cite{djelouah2019neural,yilmaz2021end}, it is non-trivial to apply the weighted mechanism when only a single reference frame is available. In this paper, only by producing multi-scale features from a single reference frame, can we make weighted prediction effective in this scenario.
\\[4pt]
\textbf{How to handle diverse motions?} The high-resolution features usually preserve more details, but they may produce residuals that are hard to compress. Sometimes low-resolution features are easier to be exploited as the context for prediction. For example, in cases of zoom-in and zoom-out, we require features at different scales to accurately predict the resolution-changing objects. For videos of flipping and rotation, features across different scales will jointly contribute to a better prediction result.  Low-resolution features can also deliver smooth prediction results in cases of explosion and fast motion with serious motion blur. We provide some examples later in Section \ref{section-analysis}.
\\[4pt]
\textbf{Differences with previous works.}
The superiority of our proposed cross-scale prediction is obvious compared with previous works. 
In particular, our method differs a lot from SSF \cite{agustsson2020scale}. 
First, our motivation is to handle diverse motions, including the uncertain motion and many other motion cases. But SSF can only blur in the uncertain motion areas such as explosion. Second, SSF uses pre-defined Gaussian blurring kernels, inflexible in its stationary blurring patterns. Here we learn multi-scale reference sources, largely extending the flexibility and accuracy of inter prediction. Third, the cross-scale prediction is designed with an advanced computing mechanism, \textit{i.e.}, weighted prediction. Unlike SSF that only selects one scale or interpolates between two scales, the mechanism of weighted prediction is basically more effective by explicitly decoding out cross-scale weight maps. This is also our essential difference with the multi-scale prediction in NVC \cite{liu2021neural}. 
In addition, the multi-head warping used here is similar to the channel grouping operation in \cite{hu2021fvc}. But we apply a linear projection layer to group the reference features. The transmitted cross-scale flows are inspired by deformable convolution \cite{dai2017deformable,zhu2019deformable}, especially multi-scale deformable convolution \cite{zhu2020deformable}. 

\subsection{Detailed Network Structures}
\label{section_3_3}
In Section \ref{section_3_1} and \ref{section_3_2}, we introduce our P-frame compression model by describing the scales of some important features. Here we add some details about the network structures. Each down/up-sampling layer is enhanced by three resblocks, the same as \cite{feng2020learned,hu2021fvc,liu2021neural}. Three are three autoencoders in our video codec, one for I-frame compression, one for motion compression and one for residual compression. Each autoencoder is enhanced by resblocks as well. The feature extractor ($X_{t}$$\rightarrow$$F_{t}$) and the final upsampling module ($\hat{F}_t$$\rightarrow$$\hat{X}_t$) consist of one down/up-sampling layer with resblocks. To compute the feature-level residual \cite{feng2020learned}, the current feature $F_{t}$ and the predicted feature $\bar{F}_{t}$ are aligned by resblocks before generating the residual $R_t$. More specific details are provided in Appendix A, including the structures of the residual block and the entropy model.

\subsection{Training and Quantization Strategy }

\label{section_3_4}

Training an entire neural video codec is a systematical task. End-to-end optimization is the superiority of neural video compression, but also results in some other issues. Here, we first introduce our training strategy in detail and then propose a multi-stage quantization strategy.
\\[5pt]
\noindent \textbf{Training Strategy} 
\\[2pt]
Our neural video codec contains an I-frame model and a P-frame model, both of which are pre-trained in advance.
To shorten the period of the initial training of the P-frame model, the target frame is first predicted with lossless reference frame. After pretraining, the I-frame and P-frame models are optimized jointly with the average rate-distortion loss. Specifically, if the training GoP size is T, then the total loss for joint training at this stage can be formulated as 
\begin{equation}
\begin{aligned}
\mathcal{L} = \mathcal{R} + \lambda \cdot \mathcal{D} 
= \mathcal{R}_{I} + \lambda \cdot \mathcal{D}_{I} + \sum_{t=1}^{T-1} ( \mathcal{R}_t + \lambda \cdot \mathcal{D}_t ).
\end{aligned}
\label{equation3}
\end{equation}
Here, $R_I$-$D_I$ and $R_t$-$D_t$ represent the rate-distortion of the I-frame and the $t$-th P-frame.
Since now the new P-frame is predicted by the previous lossy reconstructed frame, the reconstruction error will accumulate, which is known as the issue of temporal error propagation. To mitigate this issue, we could train with video sequences as long as what we evaluate on. 
However, this is infeasible in practice due to the huge GPU-memory cost. Therefore, the joint training period is further divided into several stages. We first perform joint optimization with $T=3$ and then increase $T$ to 5 to finetune the entire model. 
\\[5pt]
\noindent \textbf{Soft-then-Hard Quantization for Training Video Codec}
\\[2pt]
Since the gradient of quantization is zero almost everywhere, it makes the standard back-propagation inapplicable. A popular solution is to apply additive uniform noise (AUN) to approximate rounding during training \cite{balle2016end}, which will lead to the train-test mismatch issue \cite{agustsson2020universally,guo2021soft}. The work of \cite{guo2021soft} proposes a simple yet effective soft-then-hard (STH) strategy for training image compression models, which first optimizes with AUN and then conducts ex-post tuning. At the tuning stage, the analysis encoder $g_a$ and the hyper encoder $h_a$ are fixed. Meanwhile, the synthesis decoder $g_s$ and the entropy decoder $h_s$ are tuned with the hard-quantized variables. As a result, the decoder and the entropy model can be optimized by minimizing the true rate-distortion value.  Unlike straight-through estimator (STE) \cite{bengio2013estimating} that cannot be used for entropy estimation, STH can close the train-test mismatch in both the synthesis decoder and entropy decoder.

For video compression, since the P-frame compression model is recurrently used, the reconstruction of the previous P-frame will affect the quantized latent variables of the next P-frame. Therefore, the soft-then-hard strategy \cite{guo2021soft} designed for image codec cannot be directly applied here. We propose to gradually close the train-test mismatch of additive uniform noise, as the video-version soft-then-hard, which first performs I-frame STH to solve the I-frame train-test mismatch and then performs P-frame STH to solve the P-frame train-test mismatch. The specific process is illustrated in Table \ref{table1}. Note that at the stage of P-frame STH (stage \ding{177} in Table \ref{table1}), some of the P-frame compression modules are fixed (motion encoder, motion decoder and residual encoder). We only finetune the residual decoder and the residual entropy decoder with hard quantized latent variables. The P-frame STH strategy can help to improve the reconstruction quality of final decoded images and reduce the rate consumed for residual compression.
This multi-stage quantization strategy improves the RD performance obviously as shown in the experimental section.
\begin{table}[t]
\centering
\footnotesize

\renewcommand{\arraystretch}{1.2}
\begin{tabular}{|c|c|c|c|c|c|}
    \hline
      Training Stage & Fixed Components & Quantization Settings \\
    \hline
	\ding{172}  \makecell[c]{I-frame \\ pre-training} & None  & AUN \\
    \hline
	\ding{173}  \makecell[c]{P-frame \\ pre-training} & None & AUN \\
    \hline
	\ding{174}  \makecell[c]{Joint training \\ (T=3)} & None & AUN \\
    \hline
	\ding{175} \makecell[c]{Joint training \\ (T=5) \\ + modulated loss} & None & AUN \\
    \hline
	\ding{176} \makecell[c]{Joint training \\ (T=5) \\ I-frame STH} & \makecell[c]{I-frame $g_a$ and $h_a$} & \makecell[c]{Hard for I-frame $g_s, h_s$  \\ AUN for P-frame model} \\
    \hline
	\ding{177} \makecell[c]{Joint training \\ (T=5) \\ P-frame STH} & \makecell[c]{I-frame model \\ P-frame $g_a, h_a$} & \makecell[c]{Always hard quantization \\ Finetune P-frame $g_s, h_s$} \\
    \hline
\end{tabular}
\caption{The multi-stage quantization and training strategy. \textit{AUN} refers to additive uniform noise \cite{balle2016end}. \textit{Hard} means hard quantization without gradient passing to encoder (encoder is fixed). Stage \ding{175} and \ding{176} can actually be merged.
\label{table1}}
\end{table}
\\[3pt]
\noindent \textbf{Modulated Loss}
\\[2pt]
The modulated loss function assigns larger $\lambda$ value for the later P frames. The goal of this loss is to balance the reconstruction quality of frames in one GoP, the concept of which has been implemented in \cite{rippel2021elf,mentzer2021towards}. With its help, the later P frame will be reconstructed with better quality, mitigating temporal error propagation.
We adopt this modulated loss when $T=5$ (stage\ding{175}\ding{176}\ding{177} in Table \ref{table1}): 
\begin{equation}
\begin{aligned}
\mathcal{L}  = \mathcal{R}_{I} + \lambda \cdot \mathcal{D}_{I} + \sum_{t=1}^{T-1} ( \mathcal{R}_t + \mu_t  \cdot \lambda \cdot \mathcal{D}_t ).
\end{aligned}
\label{equation4}
\end{equation}
The $t$-th P-frame is assigned with larger distortion constraint, controlled by a monotonously increasing coefficient $\mu_t$. We set $\mu_1=1$ and $\mu_{t+1} - \mu_{t}=0.2$ in our experiments.

\section{Experiments}

\begin{figure*}[t]
 \centering
 \begin{subfigure}{0.245\linewidth}
\caption*{Previous frame}
\includegraphics[scale=0.064, clip, trim=0cm 0cm 0cm 0cm]{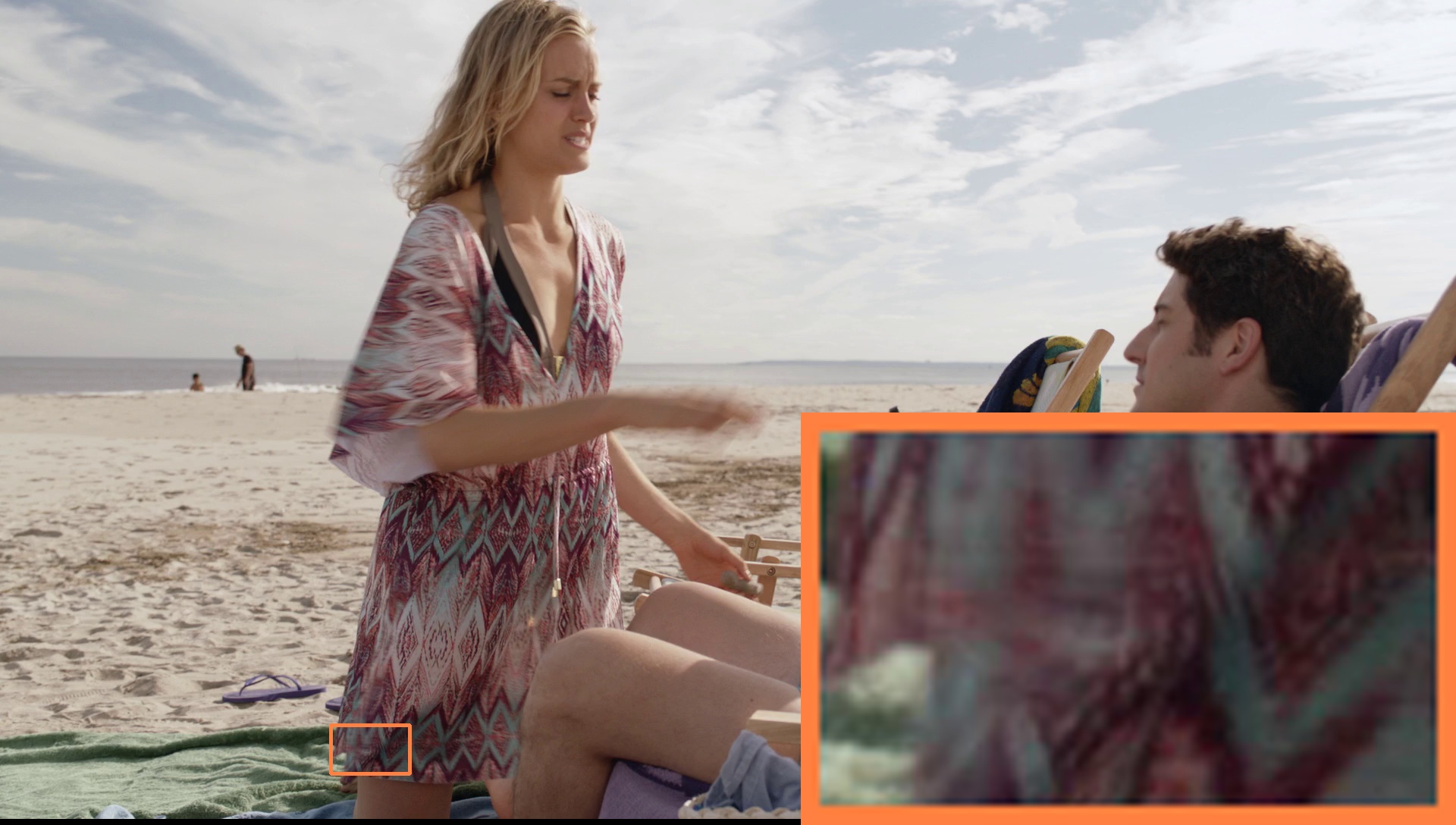}
 \end{subfigure}
 \begin{subfigure}{0.245\linewidth}
\caption*{A displacement field at scale-1}
\includegraphics[scale=0.128, clip, trim=0cm 0cm 0cm 0cm]{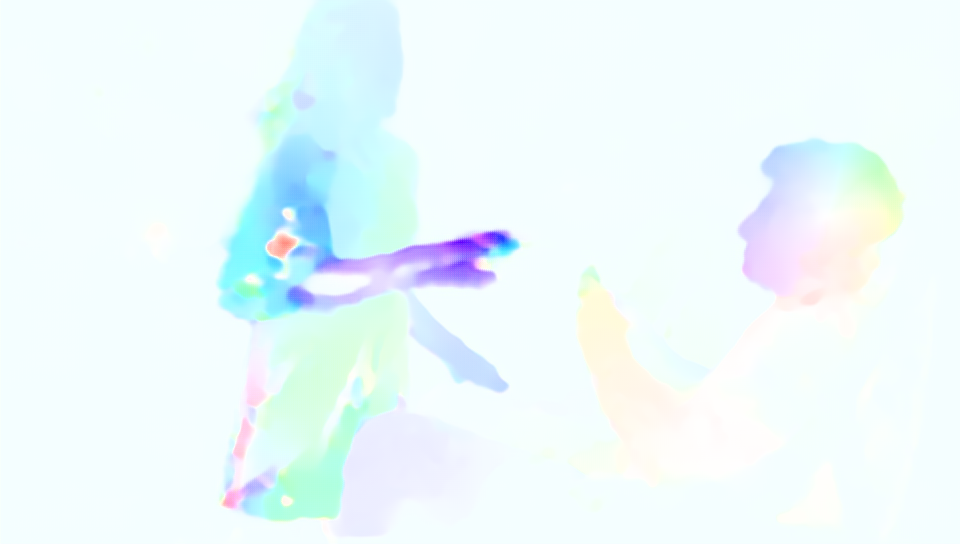}
 \end{subfigure}
 \begin{subfigure}{0.245\linewidth}
\caption*{A displacement field at scale-2}
\includegraphics[scale=0.128, clip, trim=0cm 0cm 0cm 0cm]{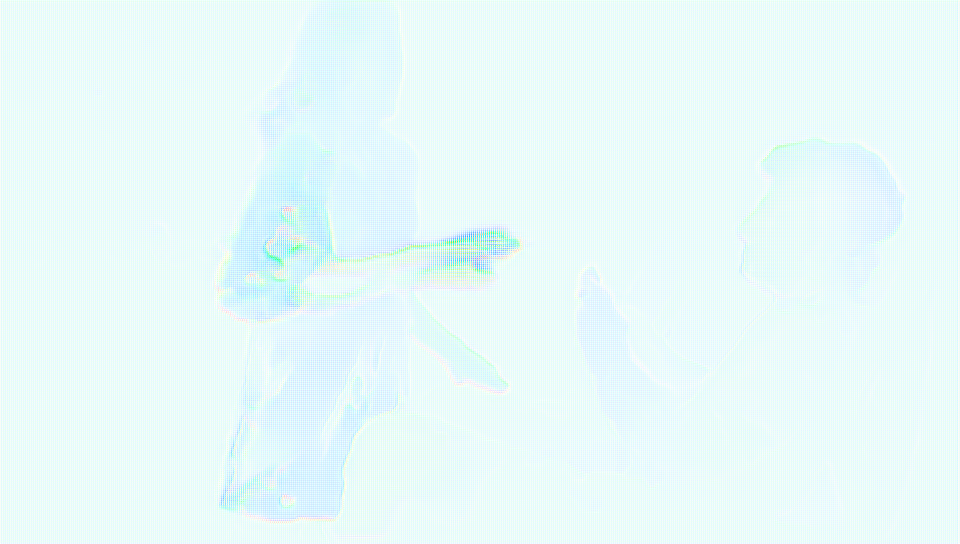}
 \end{subfigure}
 \begin{subfigure}{0.245\linewidth}
\caption*{A displacement field at scale-3}
\includegraphics[scale=0.128, clip, trim=0cm 0cm 0cm 0cm]{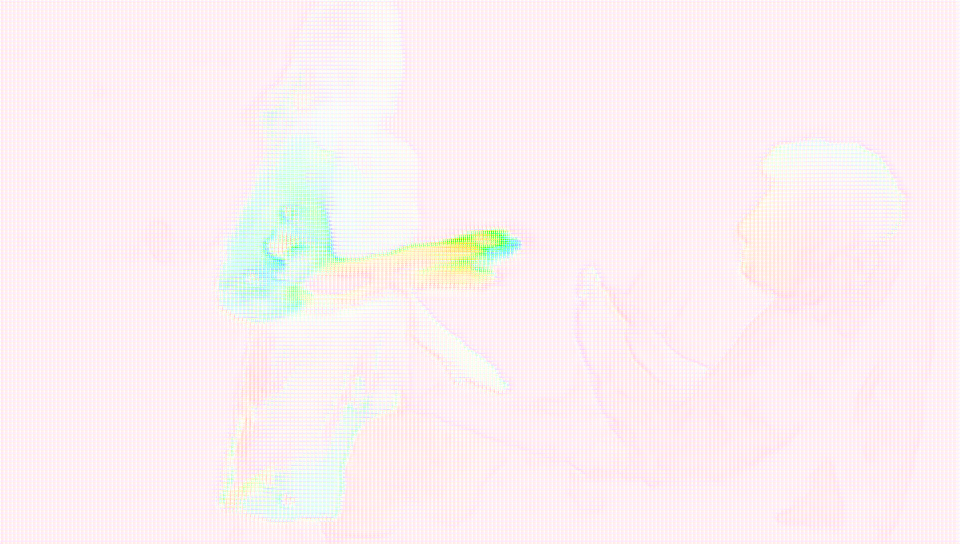}
 \end{subfigure}
\\[15pt]
  \begin{subfigure}{0.245\linewidth}
\caption*{Target frame}
\includegraphics[scale=0.064, clip, trim=0cm 0cm 0cm 0cm]{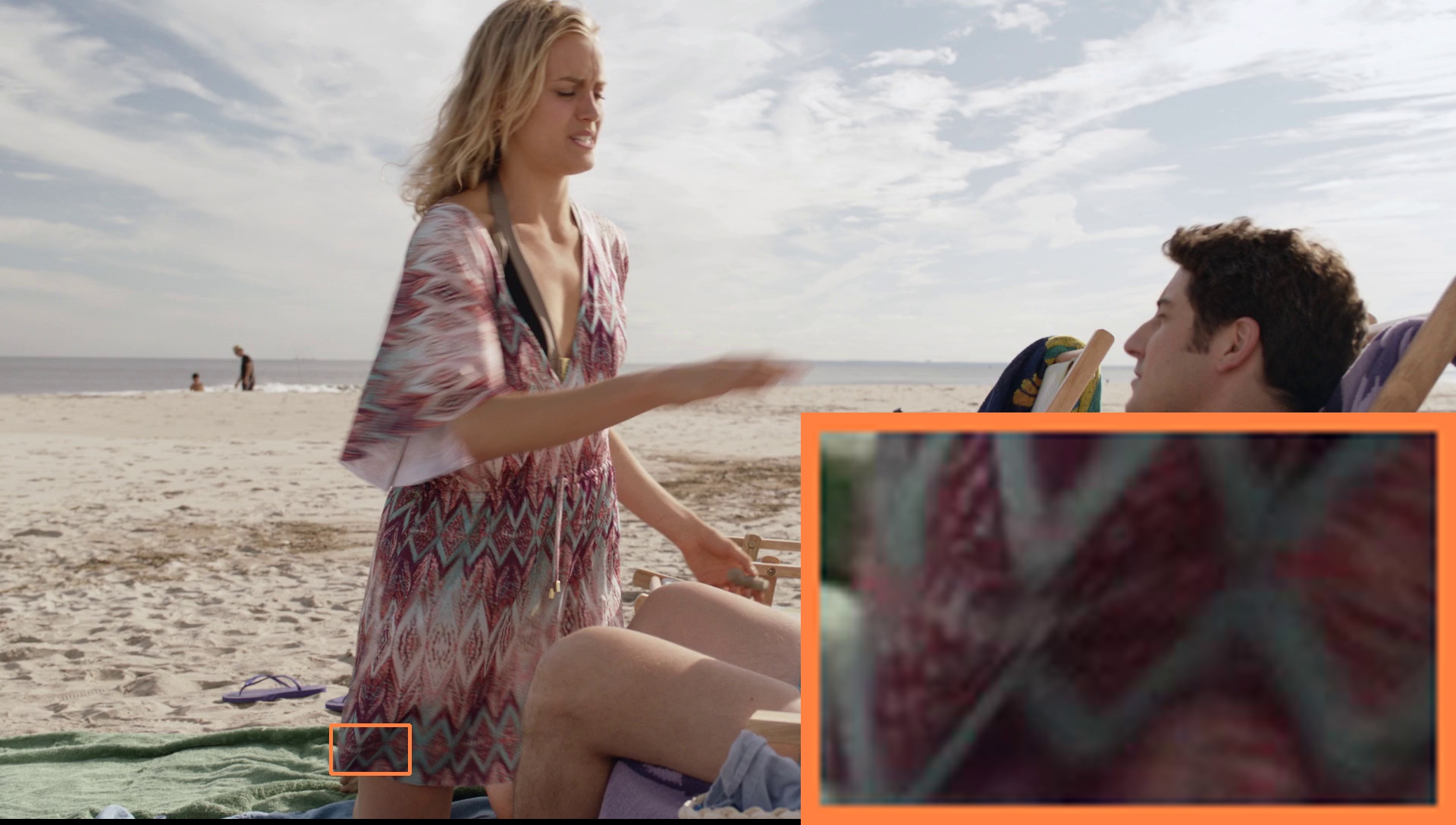}
 \end{subfigure}
 \begin{subfigure}{0.245\linewidth}
\caption*{Total weight map at scale-1}
\includegraphics[scale=0.128, clip, trim=0cm 0cm 0cm 0cm]{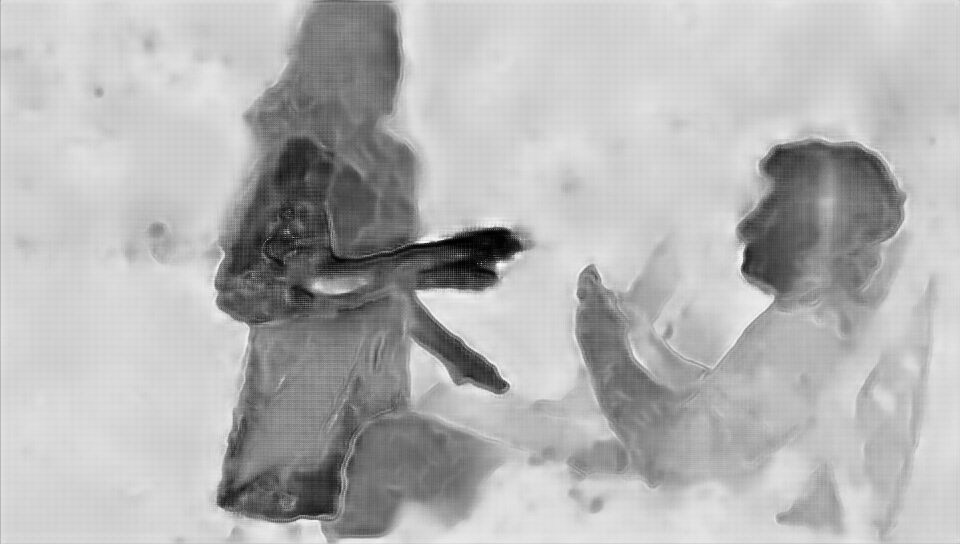}
 \end{subfigure}
 \begin{subfigure}{0.245\linewidth}
\caption*{Total weight map at scale-2}
\includegraphics[scale=0.128, clip, trim=0cm 0cm 0cm 0cm]{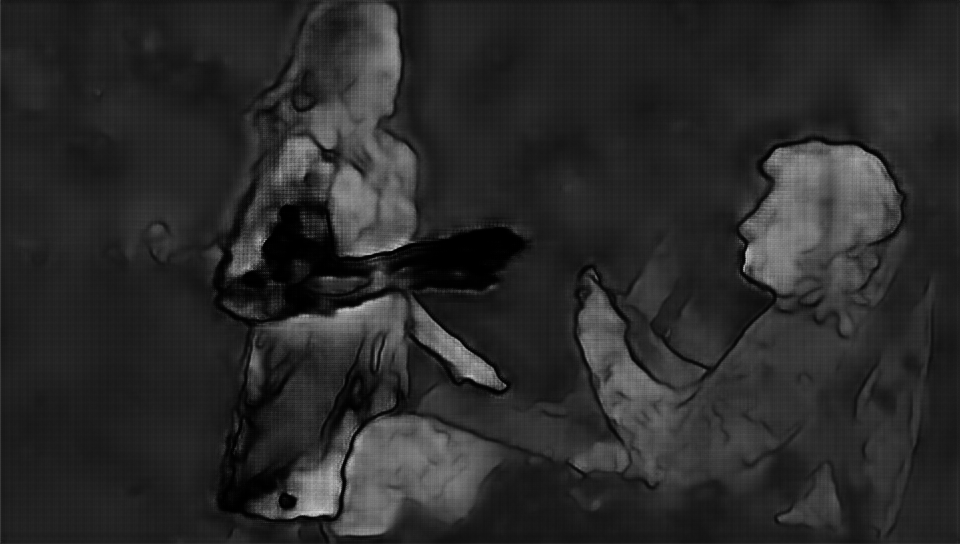}
 \end{subfigure}
\begin{subfigure}{0.245\linewidth}
\caption*{Total weight map at scale-3}
\includegraphics[scale=0.128, clip, trim=0cm 0cm 0cm 0cm]{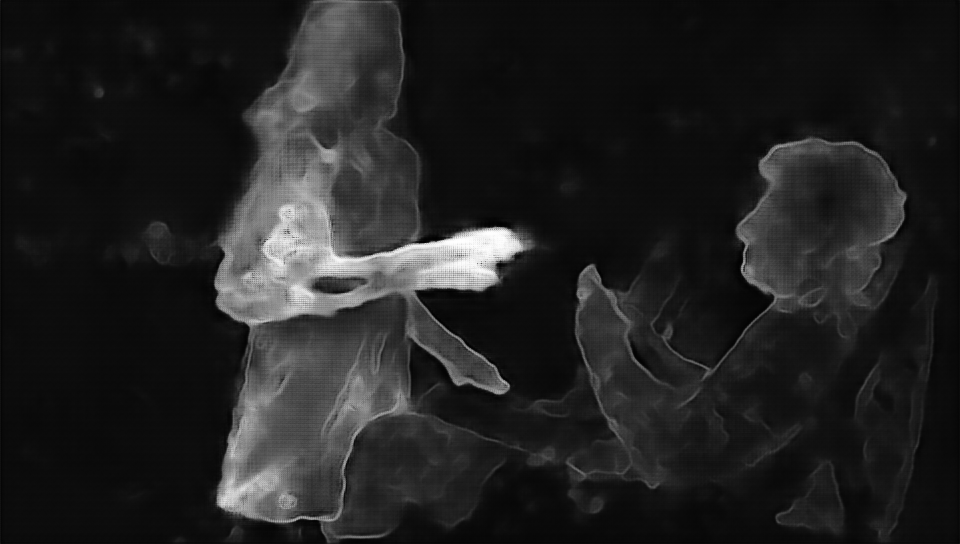}
 \end{subfigure}
\\[15pt]
  \begin{subfigure}{0.245\linewidth}
\caption*{Final reconstruction}
\includegraphics[scale=0.064, clip, trim=0cm 0cm 0cm 0cm]{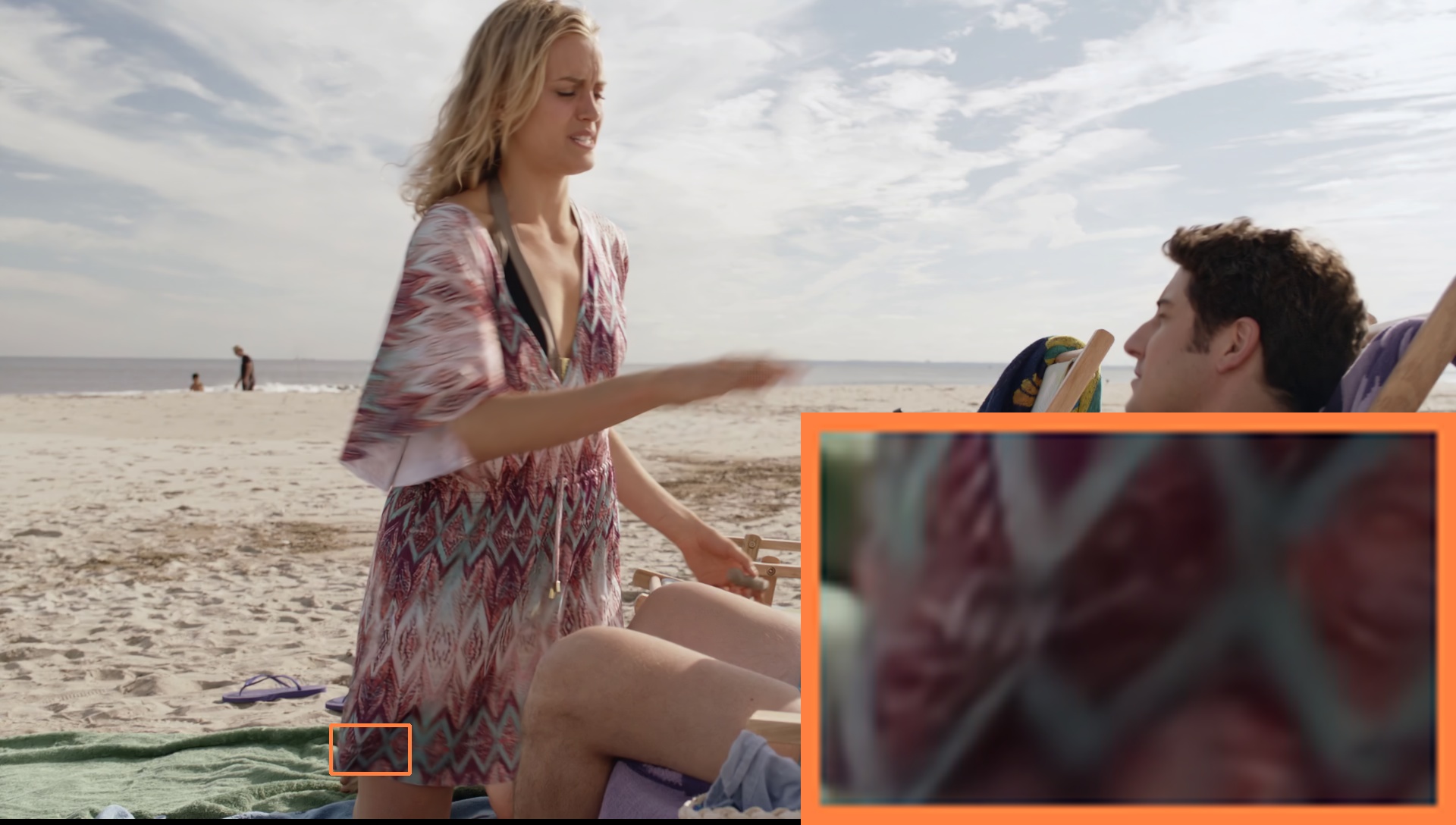}
 \end{subfigure}
 \begin{subfigure}{0.245\linewidth}
\caption*{Residual}
\includegraphics[scale=0.064, clip, trim=0cm 0cm 0cm 0cm]{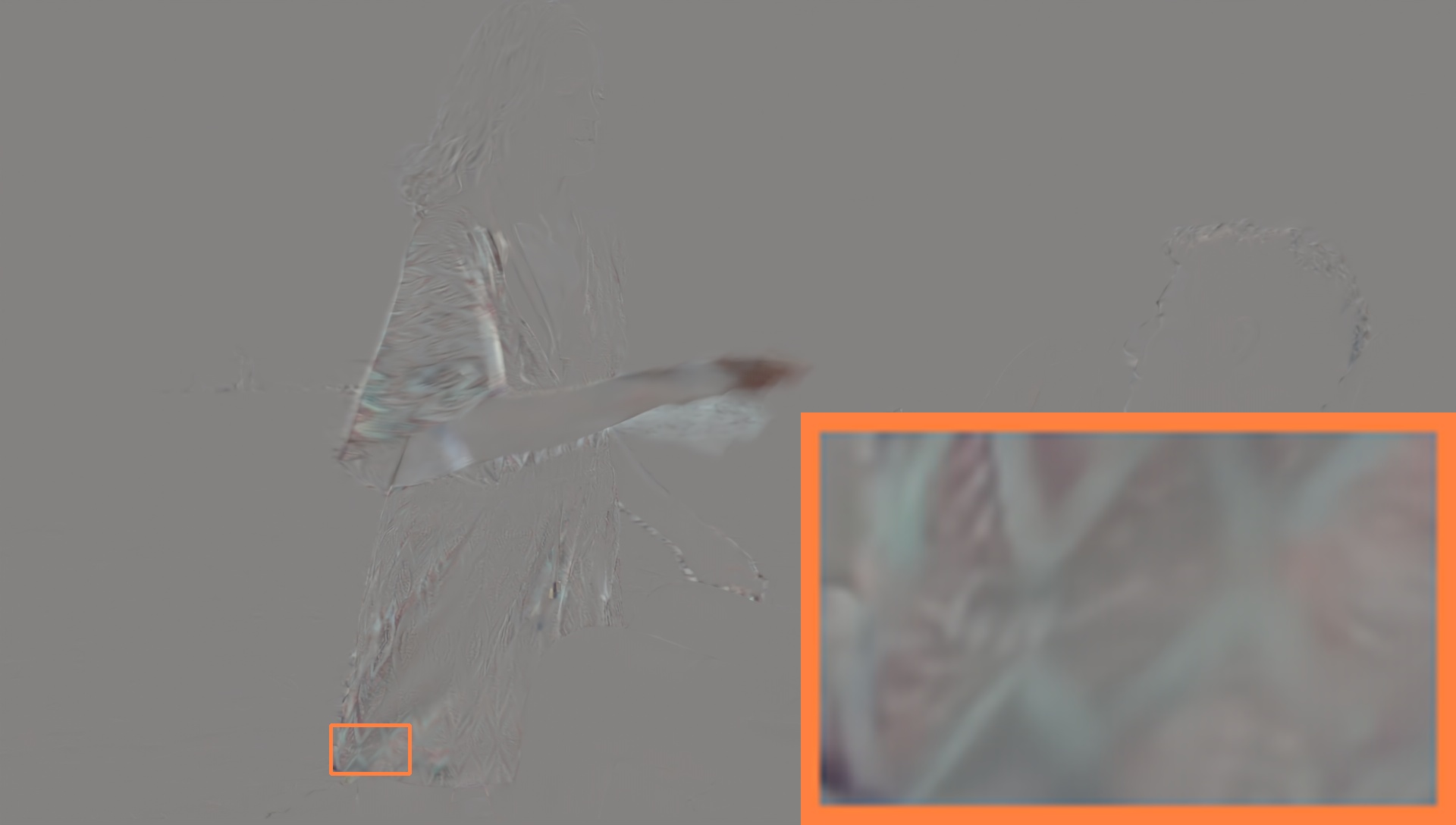}
 \end{subfigure}
 \begin{subfigure}{0.245\linewidth}
\caption*{Predicted frame across three scales}
\includegraphics[scale=0.064, clip, trim=0cm 0cm 0cm 0cm]{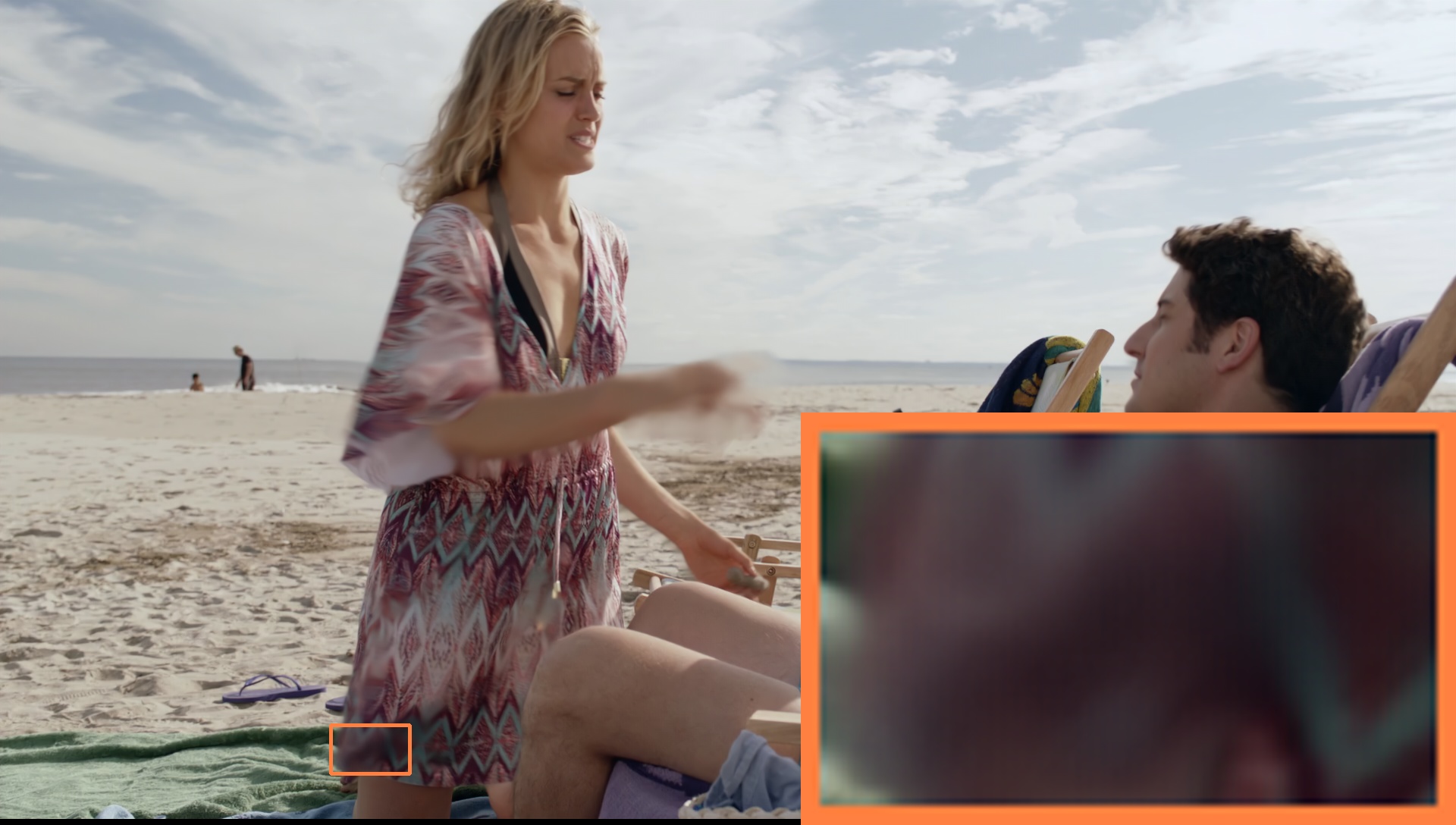}
 \end{subfigure}
\begin{subfigure}{0.245\linewidth}
\caption*{Predicted frame only with scale-1}
\includegraphics[scale=0.064, clip, trim=0cm 0cm 0cm 0cm]{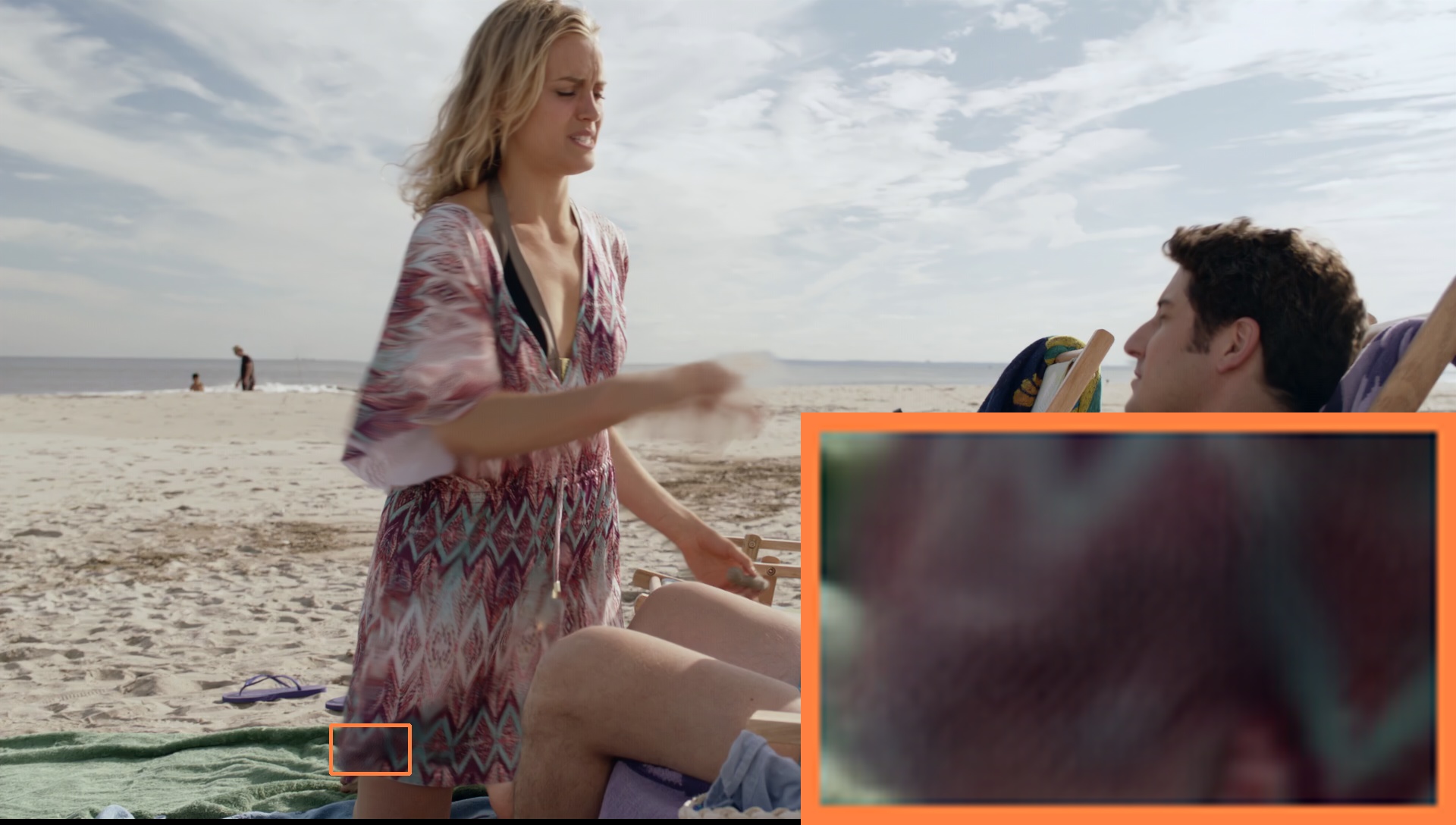}
 \end{subfigure}
 \caption{Visualizations of two consecutive frames, with corresponding cross-scale flows and cross-scale weight maps. Regarding the weight map, white color means the weight value is one and the black color means zero weight. It is found that prediction across scales can synthesize a fine prediction result which can better handle various motion cases. Noticable noise can be found if predicting only with the high-resolution scale (scale-1). Zoom-in the last picture to check the unexpected prediction noise at scale-1.}
\label{figure4}
\end{figure*}

\noindent \textbf{Datasets.} Our efficient neural video codec (ENVC) is trained on Vimeo dataset \cite{xue2019video}, which contains 89,800 video clips with the resolution of $448\times256$. Video frames are randomly cropped into $192 \times 192$ patches during training. We evaluate the performance of our proposed ENVC on multiple benchmark datasets including the UVG dataset \cite{mercat2020uvg}, HEVC test sequences \cite{sullivan2012overview} and MCL-JCV \cite{wang2016mcl}. The UVG dataset contains seven 1080p video sequences, with 3900 frames in total. The HEVC test sequences (Class B,C,D,E) contain videos with resolutions from $416\times240$ to $1920\times 1080$. The MCL-JCV dataset \cite{wang2016mcl} is another common benchmark dataset which contains 30 video sequences in 1080p.
\\[3pt]
\noindent \textbf{Implementation details.} Section \ref{section_3_4} introduces our training strategy. 
Minibatch is always set as 8 during training and we use Adam optimizer \cite{kingma2014adam}. Training the complete ENVC requires 10 days. We use two RTX 2080TI GPUs at the joint training stages.
\\[3pt]
\noindent \textbf{Standard baselines and compared methods.} For traditional standards, we compared with VTM-12.0 \cite{VTM} and HM-16.21 \cite{HM} with exactly the same prediction mode, \ieno, prediction with the previous single reference frame as well as test GoP size of 12. See Appendix B for the specific coding configurations of standard anchors. Note that the videos are compressed in YUV444 format with these anchors. 
For neural video coding approaches, we only compare with three representative hybrid coding approaches DVC \cite{lu2019dvc}, SSF \cite{agustsson2020scale} and FVC \cite{hu2021fvc}. The comparison with VTM is persuasive enough to show our SOTA performance.
\\[3pt]
\noindent \textbf{Other settings.} We do not apply variable-rate technique and thereby train multiple models for various bitrates. We set $\lambda$=\{512,1024,2048,3072,4096\} to optimize for MSE. When optimized for MS-SSIM \cite{wang2004image}, we set $\lambda$=\{6,14,24,36,50\} and finetune from the MSE-optimized model. In addition, two versions of ENVC are optimized for MSE, with or without autoregressive (AR) context model. ENVC with AR delivers the optimal RD performance but simultaneously requires much more decoding time compared with ENVC w/o AR.

\begin{figure*}[t]
 \centering
 \begin{subfigure}{0.328\linewidth}
\includegraphics[scale=0.265, clip, trim=0.3cm 0cm 0.2cm 1.1cm]{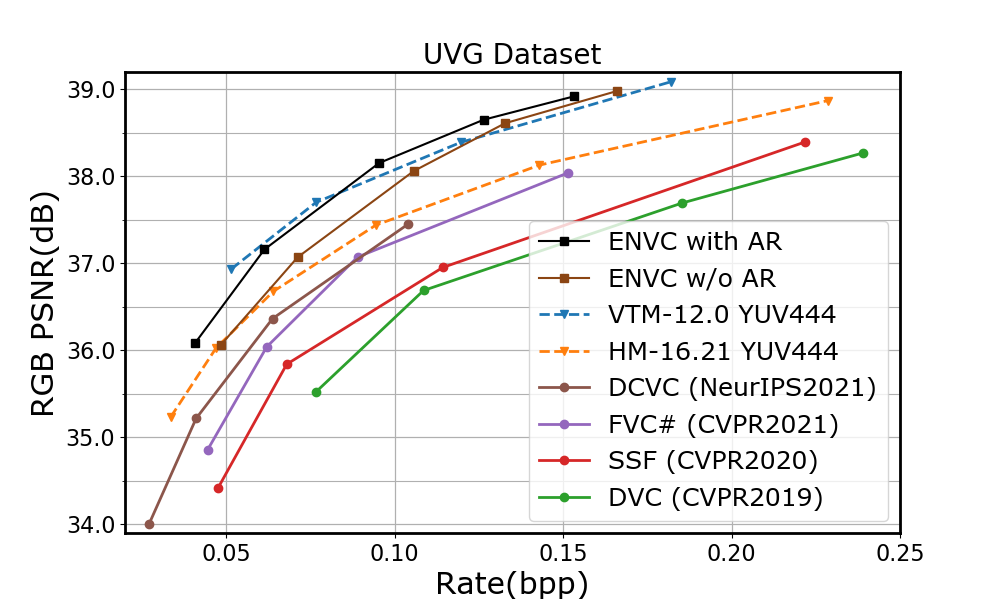}
 \end{subfigure}
 \begin{subfigure}{0.328\linewidth}
\includegraphics[scale=0.265, clip, trim=0.3cm 0cm 0.2cm 1.1cm]{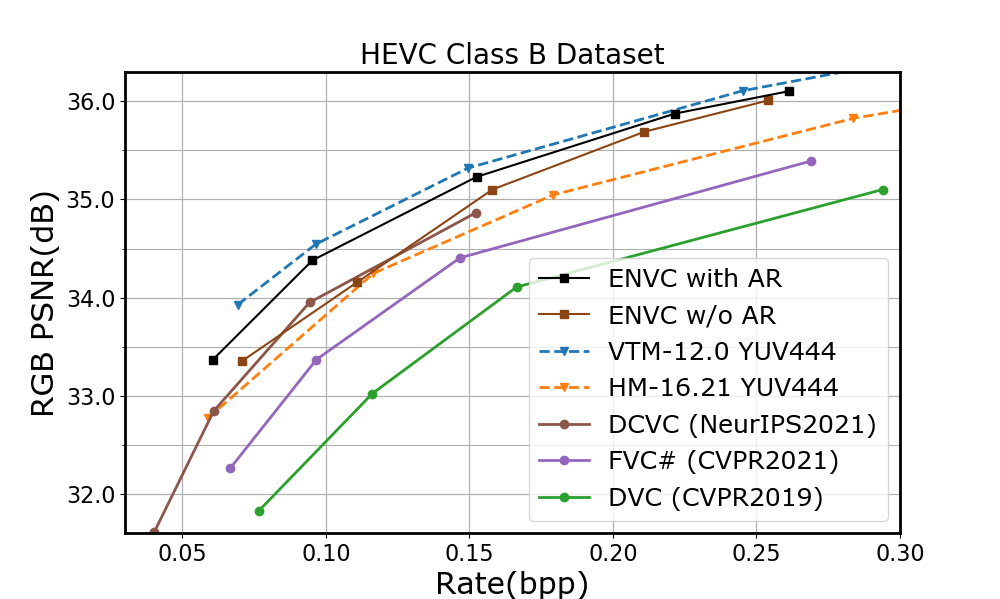}
 \end{subfigure}
 \begin{subfigure}{0.328\linewidth}
\includegraphics[scale=0.265, clip, trim=0.3cm 0cm 0.2cm 1.1cm]{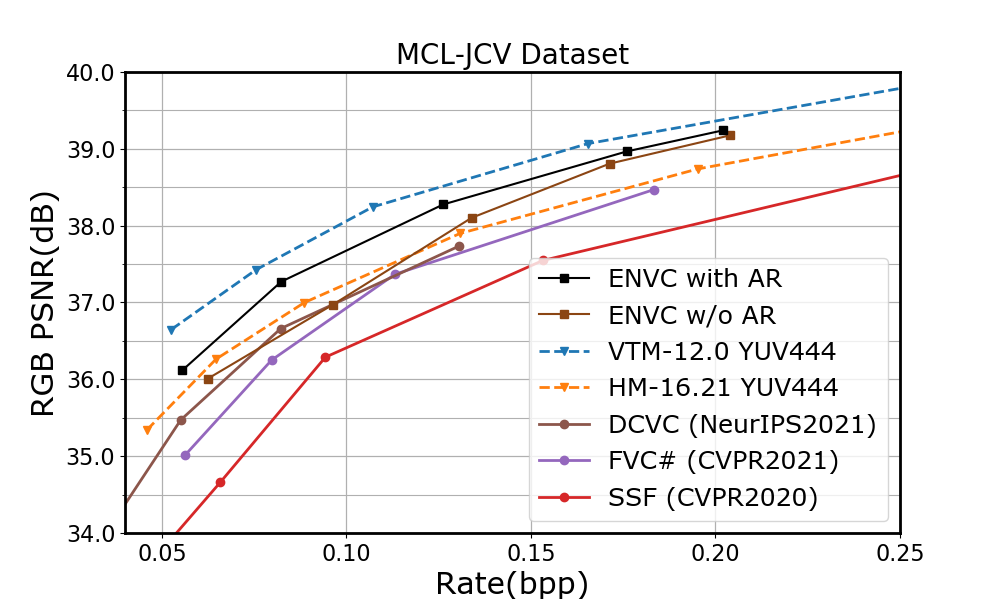}
 \end{subfigure}
\\[4pt]
 \begin{subfigure}{0.328\linewidth}
\includegraphics[scale=0.265, clip, trim=0.3cm 0cm 0.2cm 1.1cm]{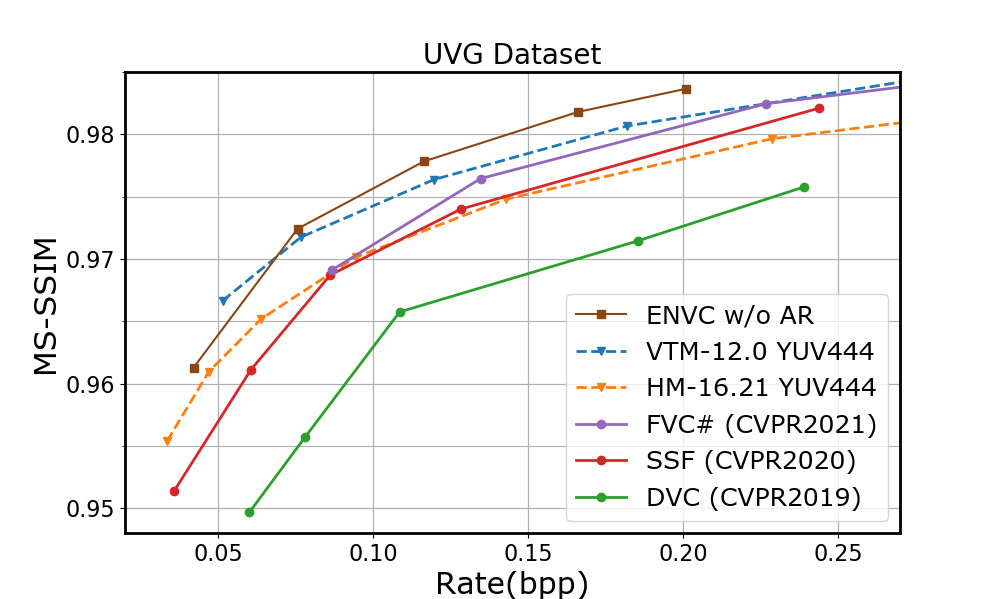}
 \end{subfigure}
 \begin{subfigure}{0.328\linewidth}
\includegraphics[scale=0.265, clip, trim=0.3cm 0cm 0.2cm 1.1cm]{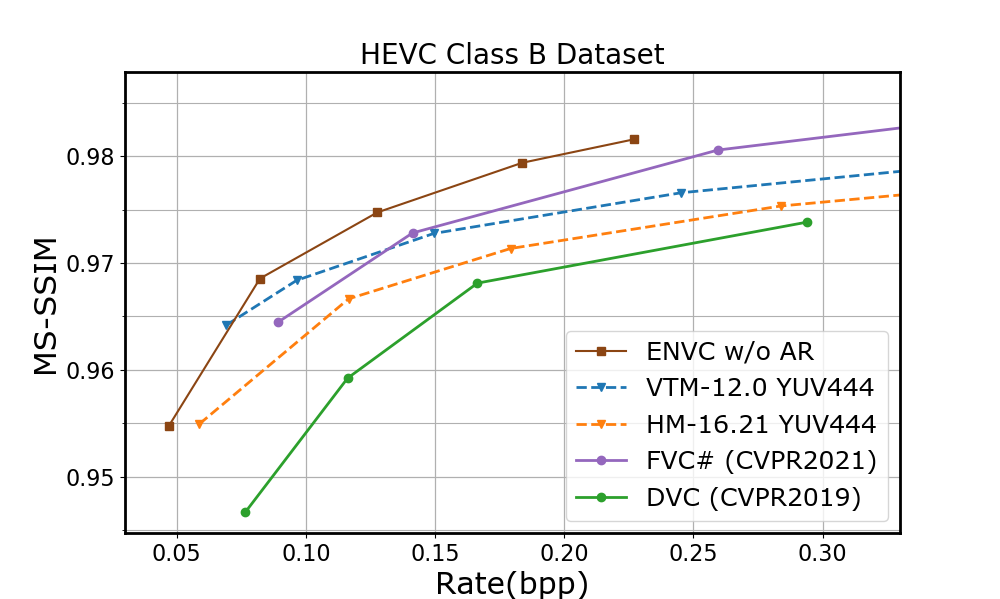}
 \end{subfigure}
 \begin{subfigure}{0.328\linewidth}
\includegraphics[scale=0.265, clip, trim=0.3cm 0cm 0.2cm 1.1cm]{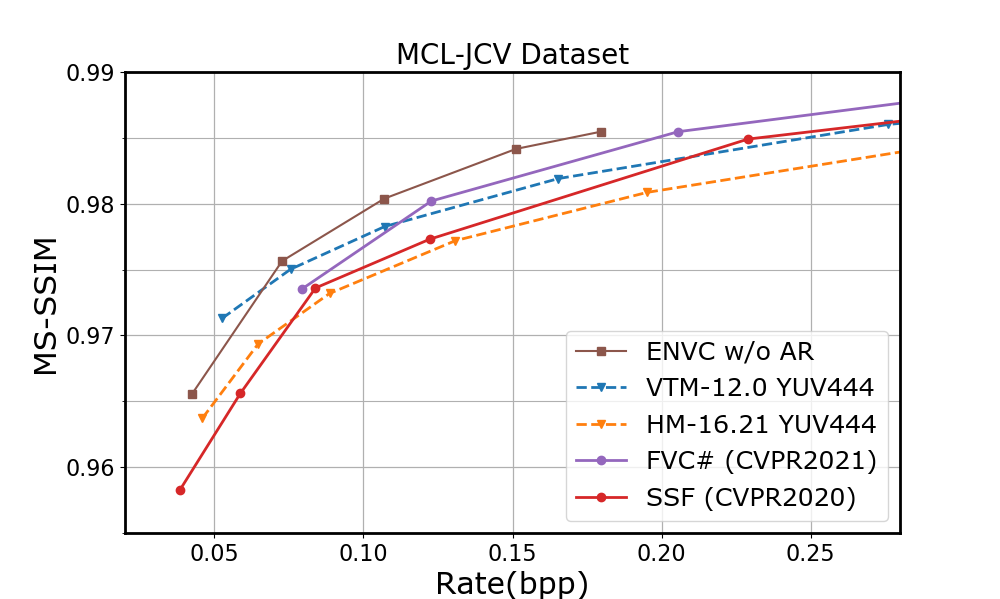}
 \end{subfigure}
 \caption{The RD-curve comparisons on UVG, HEVC Class B and MCL-JCV. We compare with previous neural video codecs according to the statistics reported in their papers. Note the performance of FVC\# \cite{hu2021fvc} is strengthened by a multiple-reference fusion module (marked by \#), but still has a large gap with our ENVC. }
\label{figure5}
\end{figure*}

\begin{table*}[t]
\centering
\small
\renewcommand{\arraystretch}{1.2}
\resizebox{\textwidth}{!}{
\begin{tabular}{|c|c|cccccc|}
\hline
\multirow{1}{*}{Metric}  & \multirow{1}{*}{Method} & \multicolumn{1}{c|}{\makecell[c]{\ \ \ \ \ \ UVG \ \ \ \ \ \ \\ 1920 $\times$ 1080}} & \multicolumn{1}{c|}{\makecell[c]{HEVC Class B \\ 1920 $\times$ 1080}} & \multicolumn{1}{c|}{\makecell[c]{HEVC Class C \\ 832 $\times$ 480}} & \multicolumn{1}{c|}{\makecell[c]{HEVC Class D \\ 416 $\times$ 240}} & \multicolumn{1}{c|}{\makecell[c]{HEVC Class E \\ 1280 $\times$ 720}} & \makecell[c]{MCL-JCV \\ 1920 $\times$ 1080} \\ \hline
\multirow{4}{*}{PSNR}    & VTM-12.0                & \multicolumn{1}{c|}{0\%} & \multicolumn{1}{c|}{\textbf{0\%}}     & \multicolumn{1}{c|}{\textbf{0\%}}          & \multicolumn{1}{c|}{\textbf{0\%}}          & \multicolumn{1}{c|}{\textbf{0\%}}          &      \textbf{0\%}        \\ 
                         & ENVC with AR             & \multicolumn{1}{c|}{\textbf{- 3.04\%}}    & \multicolumn{1}{c|}{+ 6.95\%}        & \multicolumn{1}{c|}{+ 34.74\%}             & \multicolumn{1}{c|}{+ 16.00\%}             & \multicolumn{1}{c|}{+ 16.11\%}             &     + 15.50\%    \\ 
                         & ENVC w/o AR    & \multicolumn{1}{c|}{+ 16.98\%}    & \multicolumn{1}{c|}{ + 27.44\%}        & \multicolumn{1}{c|}{ + 42.46\%}         & \multicolumn{1}{c|}{ + 20.39\%}             & \multicolumn{1}{c|}{+ 32.84\%}             &   + 31.46\%    \\ 
                         & FVC\#                & \multicolumn{1}{c|}{+ 60.62\%}    & \multicolumn{1}{c|}{+ 66.31\%}        & \multicolumn{1}{c|}{+ 53.09\%}             & \multicolumn{1}{c|}{+ 52.26\%}             & \multicolumn{1}{c|}{Unknown}             &   + 53.97\% \\ \hline
\multirow{3}{*}{MS-SSIM} & VTM-12.0                & \multicolumn{1}{c|}{0\%}    & \multicolumn{1}{c|}{0\%}        & \multicolumn{1}{c|}{0\%}             & \multicolumn{1}{c|}{0\%}             & \multicolumn{1}{c|}{0\%}             &   {0\%}           \\ 
                         & ENVC w/o AR          & \multicolumn{1}{c|}{\textbf{- 9.12\%}}    & \multicolumn{1}{c|}{\textbf{- 22.31\%}}    & \multicolumn{1}{c|}{\textbf{- 17.07\%}}   & \multicolumn{1}{c|}{\textbf{- 33.93\%}}     & \multicolumn{1}{c|}{\textbf{- 20.34\%}}             &     \textbf{- 22.02\%}         \\ 
                         & FVC\#                & \multicolumn{1}{c|}{+ 22.91\%}    & \multicolumn{1}{c|}{+ 1.77\%}        & \multicolumn{1}{c|}{- 3.64\%}             & \multicolumn{1}{c|}{- 17.24\%}             & \multicolumn{1}{c|}{Unknown}             &                - 7.01\% \\ \hline
\end{tabular}}
\caption{Calculating the BD-rate savings against VTM-12.0 on multiple benchmark datasets. $+$ means more bits required. FVC\# here is enhanced by multiple-reference prediction, but the performances of VTM and our ENVC are obtained only with single reference frame. \label{table2}}
\end{table*}

\section{Performance and Analysis}

\subsection{Comparison on RD Performance}

\noindent \textbf{RD-curves.} We compare the RD curve of ENVC with traditional and other ML-based codecs in Figure \ref{figure5}. In the single-reference prediction setting, our proposed ENVC with AR model achieves RD performance comparable to that of VTM-12.0, indicating the competitive performance of a neural video codec against the latest classical codec H.266/VVC in terms of sRGB PSNR. When evaluated by MS-SSIM, which is a more perceptual-friendly metric, ENVC exceeds VTM-12.0 especially at high bitrates. We only provide the results of ENVC w/o AR when optimized for MS-SSIM, because our experiments find AR model brings only few gains with this metric. Statistically, ENVC with AR achieves only 0.4\% rate savings against no AR version on UVG dataset. Considering we loaded the MSE-optimized models and finetuned for MS-SSIM, perhaps the finetuned models were not well converged.
Since previous neural video codecs can barely compete with H.265/HEVC, our proposed ENVC outperforms DVC \cite{lu2019dvc} and SSF \cite{agustsson2020scale} and FVC\# \cite{hu2021fvc} by a large margin. Note FVC\# has been enhanced by multiple-reference prediction. In addition, we also compare with DCVC \cite{li2021deep} although DCVC is an orthogonal work to our method which leverages the prediction result as context to facilitate the residual compression. Our method focuses only on improving the prediction process, but still outperforms DCVC.
\\[4pt]
\noindent \textbf{BD-rate savings.} To quantitatively measure the RD performance of ENVC, we calculate the BD-rate savings \cite{bjontegaard2001calculation} against VTM-12.0 in Table \ref{table2}. Specifically, the performance of ENVC (with AR) exceeds VTM slightly on UVG dataset in terms of PSNR (about 3.04\% BD-rate savings). Regarding MS-SSIM, ENVC (no AR version) achieves 9.12\% savings on UVG dataset and 33.93\% savings in Class D dataset. Compared with traditional codec VTM, it is interesting to find that neural video codecs can achieve higher PSNR on high-resolution datasets (such as UVG and Class B), consistent with the results in previous works \cite{feng2021versatile,li2021deep}.

\subsection{Analyses} \label{section-analysis}

\begin{figure}[t]
 \centering
 \begin{subfigure}{1.0\columnwidth}
 \centering
\includegraphics[scale=0.5, clip, trim=5.5cm 13.3cm 5cm 19.4cm]{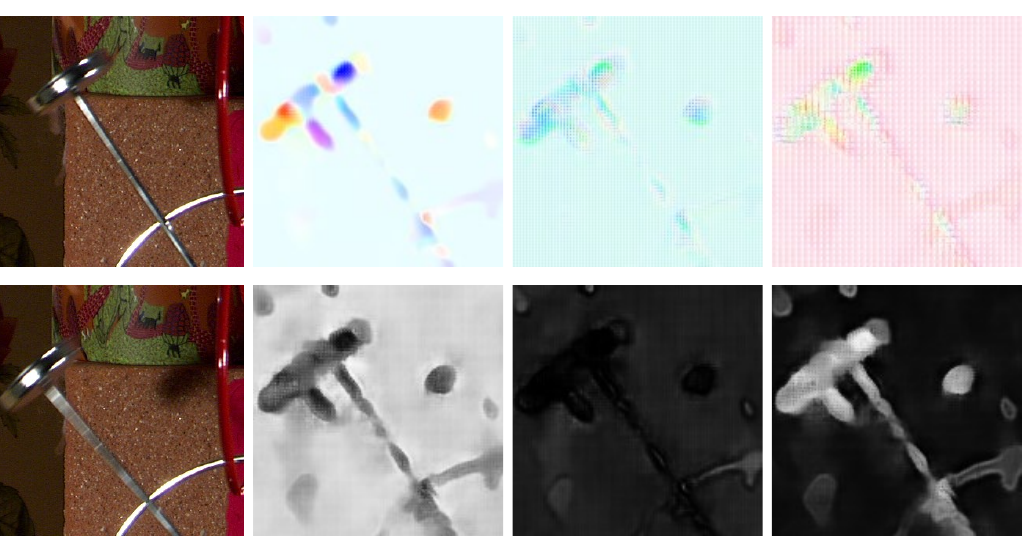}
\caption{Rotation.}
 \end{subfigure}
 \begin{subfigure}{1.0\columnwidth}
 \centering
\includegraphics[scale=0.5, clip, trim=5.5cm 31.5cm 5cm 1.2cm]{./figures/envc_various_motions}
\caption{Getting closer to the camera (an effect of zoom-in).}
 \end{subfigure}
 \caption{More visualizations of the cross-scale prediction module in the cases of rotation and zoom-in. In the first row, we visualize the reference frame, the displace field at scale-1, scale-2 and scale-3. In the second row, we visualize the target frame, the total weight map at scale-1, scale-2, scale-3 (the white color means the weight value is closer to 1). }
\label{figure6}
\end{figure}

\noindent \textbf{Visualizations.} As shown in Figure \ref{figure4}, we visualize both the cross-scale flows and weight maps. We remind our readers that although scales 1,2,3 correspond to $\frac{1}{2}$, $\frac{1}{4}$, $\frac{1}{8}$ original resolution, all the cross-scale flows and weight maps are at the scale of $\frac{H}{2} \times \frac{W}{2}$, the same as the scale of the predicted feature. There are multiple flows pointing to each scale (in our final model, N=4), and we only visualize the most important displacement field attending to each scale. It is observed that the displacement field at scale-1 is dense, which delivers a high-precision prediction. The displacement fields at scale-2 and scale-3 focus more on the region with complicated motion (such as the region of hand in Figure \ref{figure4}). Besides, 
we add all the weight maps at each scale, as the total weight map at scale-{1,2,3}. The visualizations of these weight maps support our design motivation well. For example, in Figure \ref{figure4}, the weight map at scale-3 has larger weight value in the area including the lady's hand and skirt edge, where the motion is complicated. Moreover, we can synthesize the predicted frame only with scale-1, by manually setting the weight values at scale-2 and scale-3 as zero. Now we can observe irregular noise in the zoom-in area of the last picture shown in Figure \ref{figure4}. It implies that high-resolution prediction will incur unexpected noise when motion is complicated. Prediction with low-resolution reference features can yield more compressible residuals in these areas. 
\\[4pt]
\noindent \textbf{Various video content.} The proposed cross-scale prediction module can handle various motion content. Here we visualize how this module can synthesize fine prediction results in the cases of rotation and zoom-in as shown in Figure \ref{figure6}. Sometimes low-resolution (LR) features are easier to be exploited as the context for prediction. For example, in cases of zoom-in, we require features at different scales to accurately predict the resolution-changing objects. For videos with rotation effect, the low-resolution reference features will also contribute to a better prediction result. Therefore, unlike previous Gaussian-scale prediction method \cite{agustsson2020scale} that can only blur the reference frame to generate smooth prediction results, our proposed cross-scale prediction module can adaptively learn appropriate reference features for better inter-frame prediction.

\subsection{Ablation} \label{section-ablation}

\begin{figure}[t]
 \centering
\includegraphics[scale=0.32, clip, trim=0cm 0cm 0.8cm 1.2cm]{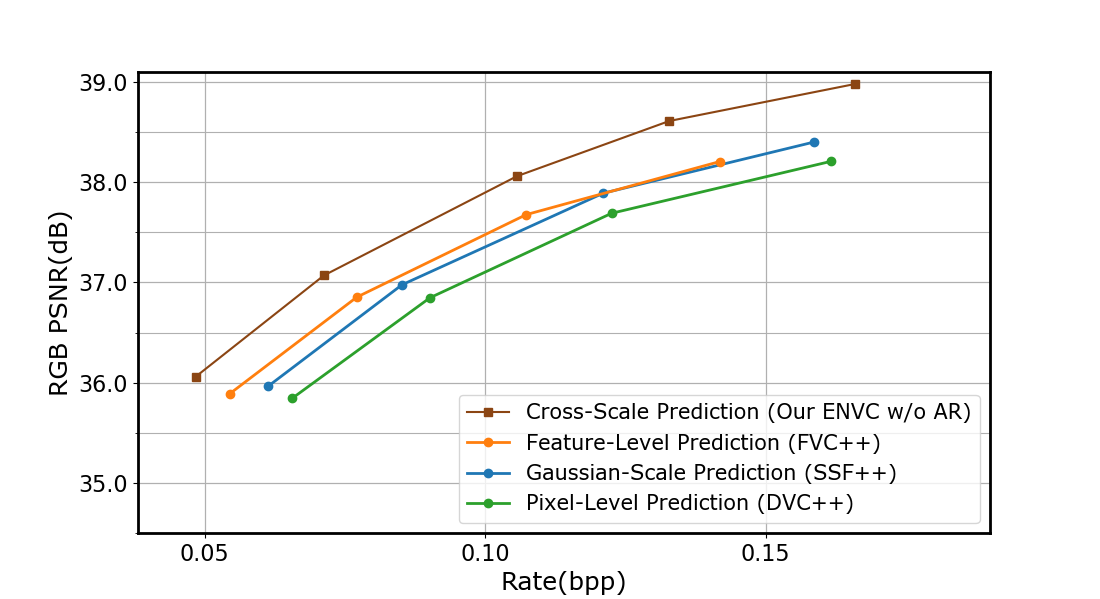}
 \caption{Ablation study of different prediction methods on UVG dataset \cite{mercat2020uvg}. Here, pixel-level residual is directly computed without any motion compensation refinement network.}
\label{figure7}
\end{figure}

\noindent \textbf{Cross-scale prediction v.s. other predictions.} Our proposed cross-scale prediction supports fine-grained adaptation to diverse motion content. Here, we conduct an ablation study to compare the cross-scale prediction with other prediction approaches, including pixel-level prediction (DVC \cite{lu2019dvc}), Gaussian-scale prediction (SSF \cite{agustsson2020scale}) and feature-level prediction (FVC \cite{hu2021fvc}). We reproduce DVC, SSF, FVC with the same tricks used in our ENVC, including computing the feature-level residual as a powerful motion compensation refinement network, applying the modulated loss function and the video-version soft-then-hard quantization strategy. As a result, the reproduced DVC, SSF and FVC have relatively higher performance than original report, thus denoted as the DVC++, SSF++, FVC++.
As shown in Figure \ref{figure7}, both Gaussian-scale and feature-level prediction outperform pixel-level prediction (DVC++). But our proposed cross-scale prediction brings another leap in RD performance, saving 27.9\% rate compared with pixel-level prediction. In addition, we can observe an increase in the percentage of motion rate shown in Table \ref{table3}, since effective prediction consumes more bits on motion information but can reduce the rate of residuals.
\begin{table}[t]
\centering
\footnotesize
\renewcommand{\arraystretch}{1.2}
\begin{tabular}{|c|c|}
\hline
\makecell[c]{Prediction method}        &  \makecell[c]{The percentage of \\ motion rate }\\ \hline
Pixel-level     &  20.23\% \\ \hline
Gaussian-scale  &  23.15\% \\ \hline
Feature-level   &  23.27\% \\ \hline
Cross-scale     &  31.11\% \\ \hline
\end{tabular}
\caption{Ablation study. The percentage of bits to encode motion information when $\lambda=1024$ (Corresponding to the models in Figure \ref{figure7}).}
\label{table3}
\end{table}
\begin{table}[t]
\centering
\footnotesize
\renewcommand{\arraystretch}{1.2}
\begin{tabular}{c c c}
    \hline
      Component & \makecell[c]{Training \\ GoP Structure} &  \makecell[c]{Estimated \\ Rate Savings} \\
    \hline
	Base Model  & I-P-P & 0\% \\
    \hline
	\makecell[c]{+ Feature Residual \cite{feng2020learned}} & I-P-P & - 13\% \\
    \hline
	\makecell[c]{More Training Frames \\ \& + Modulated Loss} & I-P-P-P-P & - 18\% \\
    \hline
	\makecell[c]{+ I-frame STH} & I-P-P-P-P & - 20\% \\
    \hline
	\makecell[c]{+ P-frame STH} & I-P-P-P-P & - 27\% \\
    \hline
	\makecell[c]{+ AR Model \\ (ENVC with AR)} & I-P-P-P-P & - 35\% \\
    \hline
\end{tabular}
\caption{Ablation study on some engineering tricks. The base model applies cross-scale prediction and computes pixel residuals (Note final ENVC computes feature residuals).
\label{table4}}
\end{table}
\\[4pt]
\noindent \textbf{Ablation on important techniques.} We also conduct a comprehensive experiment to investigate the effectiveness of other important components in our video codec, wherein we hope our engineering experience can benefit the community of neural video compression. The model with cross-scale prediction and pixel-level residual is taken as baseline. We gradually improve the RD performance by stacking techniques one by one. We calculate the BD-rate savings against the initial base mode, the results of which are presented in Table \ref{table4}. It is found that feature-level residual \cite{feng2020learned} boosts performance obviously as it covers the gain of motion compensation refinement network \cite{lu2020end}. More training frames ($T=5$) and modulated loss bring a few gains (about 5\%). Our proposed video-version soft-then-hard (I-frame STH + P-frame STH) achieves additional 9\% savings (27-18=9\%) without any complexity penalty during inference. The gain of AR model in video compression is not so obvious as in image compression \cite{minnen2018joint} (only 35-27=8\% savings here). This phenomenon can be explained because in hybrid video coding, most of the bits are spent for residual compression. Mask convolution is not well-suited to estimate the irregular context of residual.
\\[4pt]
\begin{figure}[t]
 \centering
 \includegraphics[scale=0.32, clip, trim= 0.2cm 0cm 1cm 1.5cm]{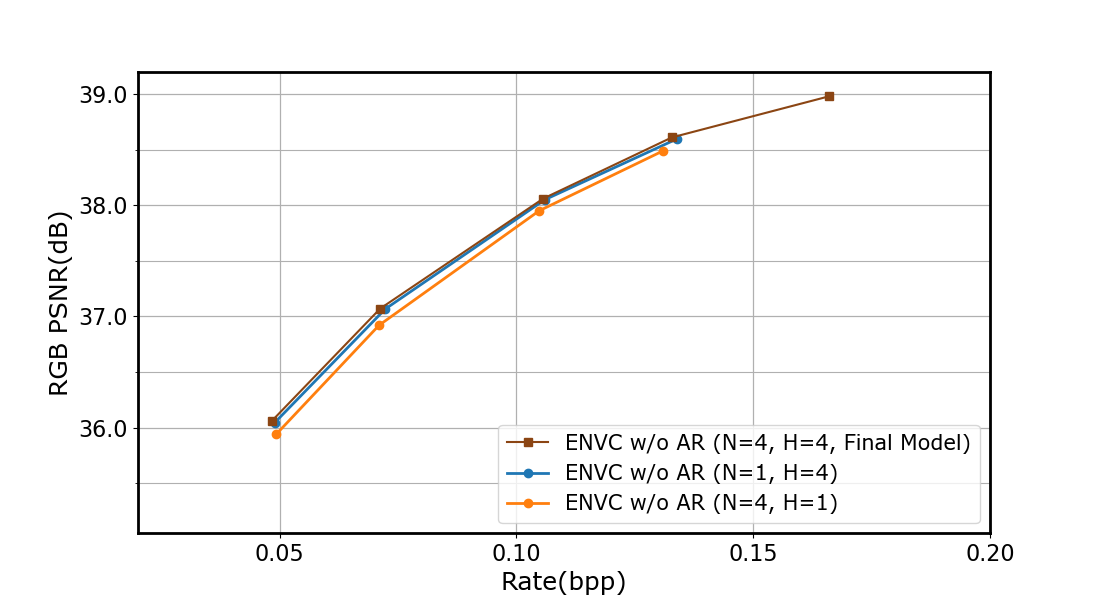}
 \caption{We study the influence of the offset number $N$ per scale and the head number $H$ on UVG dataset \cite{mercat2020uvg}.}
\label{figure8}
\end{figure}
\noindent \textbf{Ablation on some designing details.} Here, we discuss about some specific designing choices. Our final model contains a multi-head cross-scale prediction operator. The head number $H$ is set as 4 since we experimentally found that it is slightly better than a single head (\ieno, better than no re-grouping across channels). The comparisons are shown in Figure \ref{figure8}. In addition, we finally assign four offsets at each scale. Although the offset number $N$ at each scale is found to have negligible influence on final rate-distortion performance, increasing the value of $N$ is observed to stabilize and accelerate the optimization process obviously. Since the channel re-grouping process and weighted warping are performed on GPUs, the head number and the offset number at each scale have very little impact to the practical compression speed.
\\[4pt]
\begin{figure}[t]
 \centering
 \includegraphics[scale=0.32, clip, trim= 0.2cm 0cm 1cm 1.5cm]{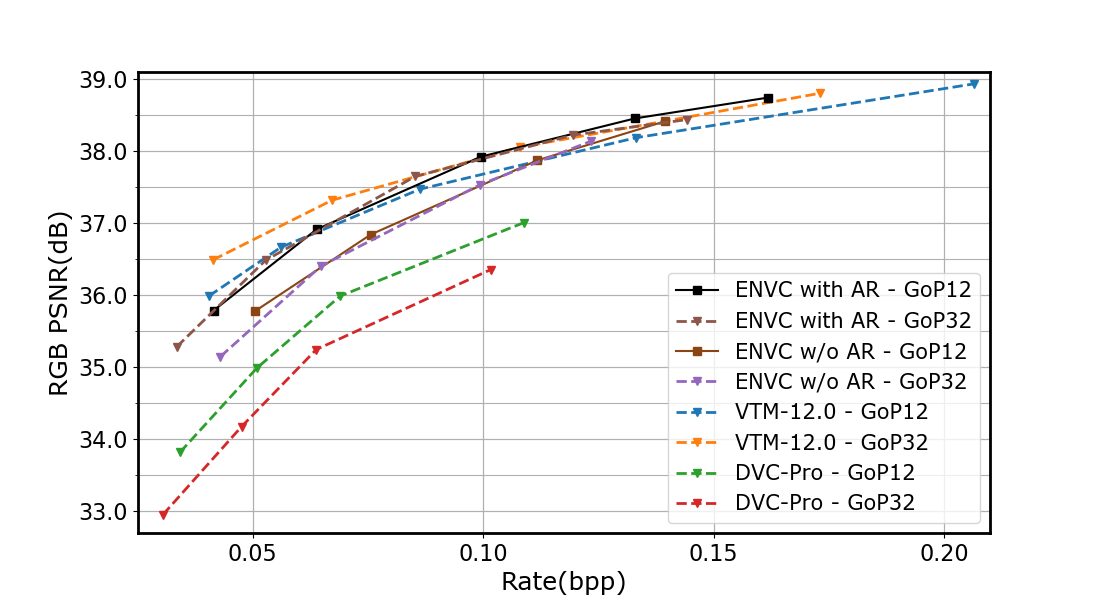}
 \caption{The RD performance when GoP size is 12 or 32. Here we evaluate different methods on UVG dataset \cite{mercat2020uvg} and only compress 96 frames for comparisons. Our method achieves close performance with GoP size as 12 or 32.}
\label{figure9}
\end{figure}
\noindent \textbf{Longer GoP size.} Most previous works \cite{lu2019dvc,agustsson2020scale,hu2021fvc} evaluate the performance of neural video codecs by setting the intra period as 12, \ieno, GoP size=12. Here, we also evaluate our ENVC with longer intra period (GoP size=32), since the proposed cross-scale prediction module has better performance for inter-frame compression. As shown in Figure \ref{figure9}, our ENVC delivers consistent rate-distortion performance when GoP size is set as either 12 or 32. However, the performance of DVC\_Pro \cite{lu2020end} will drop dramatically when increasing the GoP size from 12 to 32. It demonstrates the effectiveness of our method on practical applications with long GoP size.

\begin{table*}[t]
\centering
\small
\renewcommand{\arraystretch}{1.2}
\begin{tabular}{|c|cc|cc|}
\hline
\multirow{2}{*}{Method} & \multicolumn{2}{c|}{UVG (1920 $\times$ 1080)}         & \multicolumn{2}{c|}{HEVC Class C (832 $\times$ 480)}        \\ \cline{2-5} 
                        & \multicolumn{1}{c|}{Encoding / ms} & Decoding / ms & \multicolumn{1}{c|}{Encoding / ms} & Decoding / ms \\ \hline
ENVC with AR            & \multicolumn{1}{c|}{711 (+2910)}        &   433 (+34373)& \multicolumn{1}{c|}{142 (+510)}      &  84 (+6984)       \\ \hline
ENVC w/o AR             & \multicolumn{1}{c|}{579 (+511)}         & 337 (+567)    & \multicolumn{1}{c|}{118 (+140)}      &  70 (+165)        \\ \hline
SSF++                 & \multicolumn{1}{c|}{487 (+651)}         & 243 (+761)    & \multicolumn{1}{c|}{100 (+203)}      &  59 (+245)        \\ \hline
DVC++                   & \multicolumn{1}{c|}{586 (+645)}         & 350 (+753)    & \multicolumn{1}{c|}{125 (+195)}      &  74 (+225)        \\ \hline
\end{tabular}
\caption{Test the encoding/decoding time on RTX 2080Ti GPU. We report the average network inference time (+ practical arithmetic coding time) of P-frame. SSF \cite{agustsson2020scale} and DVC \cite{lu2019dvc} are reproduced and enhanced, as SSF++ and DVC++.\label{table5}}
\end{table*}
\noindent \textbf{About the scales.} We finally choose three scales ($\frac{1}{2},\frac{1}{4},\frac{1}{8}$ resolution) for cross-scale prediction. On the one hand, features at scale $\frac{H}{16}\times \frac{W}{16}$ have lost much context information, which cannot bring more gains. On the other hand, producing reference features with higher spatial resolution (such as in pixel space or even sub-pixel space) consumes much more GPU memory. Feature at scale $\frac{H}{2}\times \frac{W}{2}$ preserves enough context to deliver a high-precision prediction.

\begin{table*}[t]
\centering
\small
\renewcommand{\arraystretch}{1.2}
\begin{tabular}{|c|cccc|cccc|}
\hline
             & \multicolumn{4}{c|}{UVG (1920 $\times$ 1080)}                                                                                                                                                                     & \multicolumn{4}{c|}{HEVC Class C (832 $\times$ 480)}                                                                                                                                                            \\ \cline{2-9} 
             & \multicolumn{2}{c|}{Encoding}                                                                                 & \multicolumn{2}{c|}{Decoding}                                                            & \multicolumn{2}{c|}{Encoding}                                                                                & \multicolumn{2}{c|}{Decoding}                                                           \\ \cline{2-9} 
             & \multicolumn{1}{c|}{GMACs}  & \multicolumn{1}{c|}{\begin{tabular}[c]{@{}c@{}}MMACs \\ per pixel\end{tabular}} & \multicolumn{1}{c|}{GMACs}  & \begin{tabular}[c]{@{}c@{}}MMACs \\ per pixel\end{tabular} & \multicolumn{1}{c|}{GMACs} & \multicolumn{1}{c|}{\begin{tabular}[c]{@{}c@{}}MMACs \\ per pixel\end{tabular}} & \multicolumn{1}{c|}{GMACs} & \begin{tabular}[c]{@{}c@{}}MMACs \\ per pixel\end{tabular} \\ \hline
ENVC with AR & \multicolumn{1}{c|}{3438.5} & \multicolumn{1}{c|}{1.658}                                                      & \multicolumn{1}{c|}{2623.8} & 1.265                                                      & \multicolumn{1}{c|}{701.2} & \multicolumn{1}{c|}{1.756}                                                      & \multicolumn{1}{c|}{535.0} & 1.340                                                      \\ \hline
ENVC w/o AR  & \multicolumn{1}{c|}{3432.9} & \multicolumn{1}{c|}{1.655}                                                      & \multicolumn{1}{c|}{2618.1} & 1.263                                                      & \multicolumn{1}{c|}{700.0} & \multicolumn{1}{c|}{1.753}                                                      & \multicolumn{1}{c|}{533.9} & 1.337                                                      \\ \hline
SSF++        & \multicolumn{1}{c|}{2286.6} & \multicolumn{1}{c|}{1.103}                                                      & \multicolumn{1}{c|}{1468.6} & 0.708                                                      & \multicolumn{1}{c|}{466.3} & \multicolumn{1}{c|}{1.168}                                                      & \multicolumn{1}{c|}{299.5} & 0.750                                                      \\ \hline
DVC++        & \multicolumn{1}{c|}{1538.4} & \multicolumn{1}{c|}{0.742}                                                      & \multicolumn{1}{c|}{1025.3} & 0.494                                                      & \multicolumn{1}{c|}{313.7} & \multicolumn{1}{c|}{0.786}                                                      & \multicolumn{1}{c|}{209.1} & 0.524                                                      \\ \hline
\end{tabular}
\caption{Test the encoding/decoding complexity, measured by Multiply–Accumulate Operations (MACs).\label{table6}}
\end{table*}

\subsection{Complexity}

Our I-frame compression model (w/o AR) contains 12.23 M parameters and our P-frame compression model (w/o AR) contains 21.07 M parameters. Therefore, our full model is relatively light-weighted, especially the I-frame compression model when compared with previous image compression methods \cite{cheng2020learned,guo2021causal}. To evaluate the time complexity, we study on two datasets with different resolutions. The network inference time is shown in Table \ref{table5}, independent with the time of arithmetic coding which is very sensitive to practical implementations. It is found that the network inference time of ENVC w/o AR is between the enhanced SSF \cite{agustsson2020scale} and DVC \cite{lu2019dvc}. Because on the one hand, DVC is an early framework containing an explicit optical flow extractor. On the other hand, our proposed cross-scale prediction module relies on bilinear sampling, which can be computed very fast and thereby does not increase the coding time obviously. In addition, ENVC with AR model requires much more coding time, mainly spent for arithmetic coding, especially arithmetic decoding. We also report the encoding/decoding complexity, measured by Multiply–Accumulate Operations (MACs), in Table \ref{table6}. According to Table \ref{table6}, we can find the proposed ENVC requires more multiply computations but the computational cost is still in a practical level.

\section{Conclusion}

In this paper, we present an efficient neural video codec (ENVC) with a novel cross-scale prediction module. The proposed cross-scale prediction module achieves more effective motion compensation result through fine-grained adaptation to diverse motions.
Cross-scale flows are transmitted to control the prediction precisions with a reference feature pyramid. There are also cross-scale weight maps transmitted for effective weighted prediction, which is analyzed with diverse motions. We carefully build the model and adopt a new multi-stage quantization strategy. ENVC demonstrates that a neural video codec can compete with VTM regarding sRGB PSNR in the scenario of single-reference prediction following a typical hybrid coding framework. It can further be extended to other prediction scenarios in future work.

\section*{Appendix A: Network Structures}

\begin{figure}[t]
 \centering
 \begin{subfigure}{0.49\columnwidth}
 \centering
\includegraphics[scale=0.8]{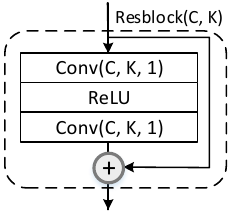}
\caption{Resblock \label{figure10a}}
 \end{subfigure}
 \begin{subfigure}{0.49\columnwidth}
 \centering
\includegraphics[scale=0.8]{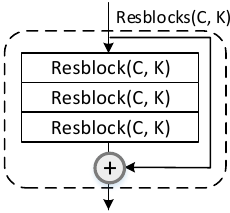}
\caption{Resblocks \label{figure10b}}
 \end{subfigure}
\\[4pt]
 \begin{subfigure}{\columnwidth}
 \centering
\includegraphics[scale=0.8]{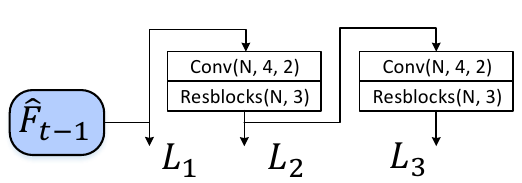}
\caption{Autoencoder \label{figure10c}}
 \end{subfigure}\\[4pt]
 \begin{subfigure}{\columnwidth}
\includegraphics[scale=0.9]{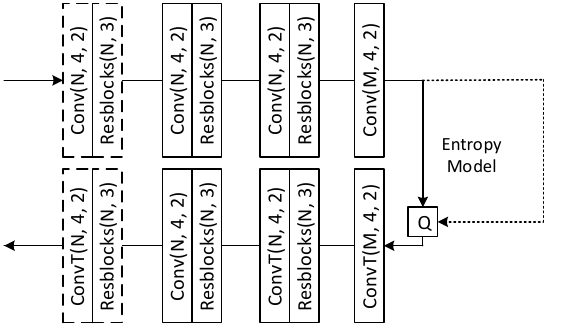}
\caption{Multi-Scale feature extractor \label{figure10d}}
 \end{subfigure}
 \caption{Specific network structures. \label{figure10}}
\end{figure}

All the enhancement modules used in our model are resblocks. We here explain how these resblocks are built. As shown in Figure \ref{figure10a}, one resblock refers to a residual connection with two convolutional layers. We stack three successive resblocks as one unit to enhance the network, which is termed as resblocks as shown in Figure \ref{figure10b}. The structures of the autoencoder used for I-frame compression are described in Figure \ref{figure10c}. We can observe that every down/up-sampling layer is followed by the unit of resblocks. Note that there are four downsampling layers in I-frame encoder. However, since we apply cross-scale feature prediction and compute the feature-level residual, there are only three downsampling layers in both the motion and the residual encoders. In addition, the structures of the multi-scale feature extractor are described in Figure \ref{figure10d}, where two downsampling layers with resblocks will transform the original reference feature from the scale of $\frac{H}{2} \times \frac{W}{2}$ to the scale of $\frac{H}{4} \times \frac{W}{4}$ and $\frac{H}{8} \times \frac{W}{8}$. We also compute the feature-level residual by applying one module of resblocks to refine the results of cross-scale prediction.

Following most previous works, we control the network capacity with two hyper parameters $N$ and $M$ as the channel number. In the I-frame compression network, we set $N=128$ and $M=160$. In the motion compression network, we set $N=64$ and $M=80$ because the rate of motion is relatively small. In the residual compression network, we set $N=128$ and $M=160$.
Regarding the entropy model, we use three kinds of entropy model, including the conventional factorized model \cite{balle2016end} and hyperprior entropy model \cite{balle2018variational} and the autoregressive entropy model \cite{minnen2018joint}. For ENVC w/o AR, we apply hyperprior model for I-frame compression and residual compression. Factorized model is applied for motion compression. For ENVC with AR, we apply autoregressive entropy model in all three compression networks. However, the entropy model in our motion compression network does not contain a hyperprior branch. That is to say, we directly use mask convolution and parameterize the motion variables as GMM distribution. We find this kind of AR model is stable and more effective for motion compression, because the context of motion information is very smooth.

\section*{Appendix B: Configurations of HM and VTM}
All the source test videos are in YUV420 format, which is also the default input format of HM and VTM. However, existing learning-based methods are optimized and evaluated in RGB color space. To make a fair comparison, we first convert the test videos to RGB images as the input of learning-based codecs, and then convert them back into YUV444 format as the input of the traditional codecs. All codecs are finally evaluated in RGB color space. The detailed coding configurations are described as follows.

\paragraph{HM-16.21}
We use the default configuration file ``encoder\_lowdelay\_P\_main.cfg'' of HM-16.21 and encode the test videos with the following command:
\begin{equation}
\begin{aligned}
&\rm TAppEncoderStatic \ 
\rm \verb|-|{c} \ [CFG] \ \verb|-|{i} \ [IN].yuv \ \verb|-|{b} \ [OUT].bin\\ 
&\rm \verb|-|{o} \ [OUT].yuv \ \verb|-|{wdt} \ [W] \ \verb|-|{hgt} \ [H] \ \verb|-|{fr} \ [FPS] \ \verb|-|{f} \ [N] \\ 
&\rm \verb|-|{q} \ [QP] \ \verb|--|{IntraPeriod\verb|=|12} \ \verb|--|{Profile\verb|=|main\_444} \\
&\rm \verb|--|{InputChromaFormat\verb|=|444} \ \verb|--|{Level\verb|=|6.1} \\ 
&\rm \verb|--|{ConformanceWindowMode\verb|=|1}
\notag       
\end{aligned}
\end{equation}
Here, \textit{N} is the number of encoded frames, which is set as 100 for HEVC datasets and 600 for UVG dataset.

\paragraph{VTM-12.0}
We use the default configuration file ``encoder\_lowdelay\_P\_vtm.cfg'' of VTM-12.0 and modified the GoP structure to fit the keyframe (I-frame) interval of 12.
We encode the test videos with the following command:
\begin{equation}
\begin{aligned}
&\rm EncoderAppStatic \ 
\rm \verb|-|{c} \ [CFG] \ \verb|-|{i} \ [IN].yuv \ \verb|-|{b} \ [OUT].bin \\
&\rm \verb|-|{o} \ [OUT].yuv \ \verb|-|{wdt} \ [W] \ \verb|-|{hgt} \ [H] \ \verb|-|{fr} \ [FPS] \ \verb|-|{f} \ [N] \\
&\rm \verb|-|{q} \ [QP] \ \verb|--|{IntraPeriod\verb|=|12} \ \verb|-|{c} \ yuv444.cfg \\
&\rm \verb|--|{InputBitDepth\verb|=|8} \ \verb|--|{OutputBitDepth\verb|=|8} \\ 
&\rm \verb|--|{InputChromaFormat\verb|=|444} \ \verb|--|{Level\verb|=|6.1} \\ 
&\rm \verb|--|{DecodingRefreshType\verb|=|2} \\
&\rm \verb|--|{ConformanceWindowMode\verb|=|1}
\notag       
\end{aligned}
\end{equation}
Here, \textit{N} is the number of encoded frames, which is set as 100 for HEVC datasets and 600 for UVG dataset.


\bibliographystyle{IEEEtran}
\bibliography{reference}

\begin{thebibliography}{10}
\providecommand{\url}[1]{#1}
\csname url@samestyle\endcsname
\providecommand{\newblock}{\relax}
\providecommand{\bibinfo}[2]{#2}
\providecommand{\BIBentrySTDinterwordspacing}{\spaceskip=0pt\relax}
\providecommand{\BIBentryALTinterwordstretchfactor}{4}
\providecommand{\BIBentryALTinterwordspacing}{\spaceskip=\fontdimen2\font plus
\BIBentryALTinterwordstretchfactor\fontdimen3\font minus
  \fontdimen4\font\relax}
\providecommand{\BIBforeignlanguage}[2]{{%
\expandafter\ifx\csname l@#1\endcsname\relax
\typeout{** WARNING: IEEEtran.bst: No hyphenation pattern has been}%
\typeout{** loaded for the language `#1'. Using the pattern for}%
\typeout{** the default language instead.}%
\else
\language=\csname l@#1\endcsname
\fi
#2}}
\providecommand{\BIBdecl}{\relax}
\BIBdecl

\bibitem{balle2018variational}
J.~Ball{\'{e}}, D.~Minnen, S.~Singh, S.~J. Hwang, and N.~Johnston,
  ``Variational image compression with a scale hyperprior,'' in \emph{6th
  International Conference on Learning Representations, {ICLR} 2018}, 2018.

\bibitem{cheng2020learned}
Z.~Cheng, H.~Sun, M.~Takeuchi, and J.~Katto, ``Learned image compression with
  discretized gaussian mixture likelihoods and attention modules,'' in
  \emph{Proceedings of the IEEE/CVF Conference on Computer Vision and Pattern
  Recognition}, 2020, pp. 7939--7948.

\bibitem{mentzer2020high}
F.~Mentzer, G.~D. Toderici, M.~Tschannen, and E.~Agustsson, ``High-fidelity
  generative image compression,'' in \emph{Advances in Neural Information
  Processing Systems}, vol.~33, 2020, pp. 11\,913--11\,924.

\bibitem{guo2021causal}
Z.~Guo, Z.~Zhang, R.~Feng, and Z.~Chen, ``Causal contextual prediction for
  learned image compression,'' \emph{IEEE Transactions on Circuits and Systems
  for Video Technology}, pp. 1--1, 2021.

\bibitem{wiegand2003overview}
T.~Wiegand, G.~J. Sullivan, G.~Bjontegaard, and A.~Luthra, ``Overview of the h.
  264/avc video coding standard,'' \emph{IEEE Transactions on circuits and
  systems for video technology}, vol.~13, no.~7, pp. 560--576, 2003.

\bibitem{sullivan2012overview}
G.~J. Sullivan, J.-R. Ohm, W.-J. Han, and T.~Wiegand, ``Overview of the high
  efficiency video coding (hevc) standard,'' \emph{IEEE Transactions on
  Circuits and Systems for Video Technology}, vol.~22, no.~12, pp. 1649--1668,
  2012.

\bibitem{bross2021overview}
B.~Bross, Y.-K. Wang, Y.~Ye, S.~Liu, J.~Chen, G.~J. Sullivan, and J.-R. Ohm,
  ``Overview of the versatile video coding (vvc) standard and its
  applications,'' \emph{IEEE Transactions on Circuits and Systems for Video
  Technology}, vol.~31, no.~10, pp. 3736--3764, 2021.

\bibitem{zhang2013background}
X.~Zhang, T.~Huang, Y.~Tian, and W.~Gao, ``Background-modeling-based adaptive
  prediction for surveillance video coding,'' \emph{IEEE Transactions on Image
  Processing}, vol.~23, no.~2, pp. 769--784, 2013.

\bibitem{zhang2018improved}
K.~Zhang, Y.-W. Chen, L.~Zhang, W.-J. Chien, and M.~Karczewicz, ``An improved
  framework of affine motion compensation in video coding,'' \emph{IEEE
  Transactions on Image Processing}, vol.~28, no.~3, pp. 1456--1469, 2018.

\bibitem{lu2019dvc}
G.~Lu, W.~Ouyang, D.~Xu, X.~Zhang, C.~Cai, and Z.~Gao, ``Dvc: An end-to-end
  deep video compression framework,'' in \emph{Proceedings of the IEEE/CVF
  Conference on Computer Vision and Pattern Recognition}, 2019, pp.
  11\,006--11\,015.

\bibitem{chen2019learning}
Z.~Chen, T.~He, X.~Jin, and F.~Wu, ``Learning for video compression,''
  \emph{IEEE Transactions on Circuits and Systems for Video Technology},
  vol.~30, no.~2, pp. 566--576, 2019.

\bibitem{habibian2019video}
A.~Habibian, T.~v. Rozendaal, J.~M. Tomczak, and T.~S. Cohen, ``Video
  compression with rate-distortion autoencoders,'' in \emph{Proceedings of the
  IEEE/CVF International Conference on Computer Vision}, 2019, pp. 7033--7042.

\bibitem{agustsson2020scale}
E.~Agustsson, D.~Minnen, N.~Johnston, J.~Balle, S.~J. Hwang, and G.~Toderici,
  ``Scale-space flow for end-to-end optimized video compression,'' in
  \emph{Proceedings of the IEEE/CVF Conference on Computer Vision and Pattern
  Recognition}, 2020, pp. 8503--8512.

\bibitem{lin2020m}
J.~Lin, D.~Liu, H.~Li, and F.~Wu, ``M-lvc: multiple frames prediction for
  learned video compression,'' in \emph{Proceedings of the IEEE/CVF Conference
  on Computer Vision and Pattern Recognition}, 2020, pp. 3546--3554.

\bibitem{liu2020conditional}
J.~Liu, S.~Wang, W.-C. Ma, M.~Shah, R.~Hu, P.~Dhawan, and R.~Urtasun,
  ``Conditional entropy coding for efficient video compression,'' in
  \emph{Computer Vision--ECCV 2020: 16th European Conference, Glasgow, UK,
  August 23--28, 2020, Proceedings, Part XVII 16}.\hskip 1em plus 0.5em minus
  0.4em\relax Springer, 2020, pp. 453--468.

\bibitem{hu2021fvc}
Z.~Hu, G.~Lu, and D.~Xu, ``Fvc: A new framework towards deep video compression
  in feature space,'' in \emph{Proceedings of the IEEE/CVF Conference on
  Computer Vision and Pattern Recognition}, 2021, pp. 1502--1511.

\bibitem{liu2021neural}
H.~Liu, M.~Lu, Z.~Ma, F.~Wang, Z.~Xie, X.~Cao, and Y.~Wang, ``Neural video
  coding using multiscale motion compensation and spatiotemporal context
  model,'' \emph{IEEE Transactions on Circuits and Systems for Video
  Technology}, vol.~31, no.~8, pp. 3182--3196, 2021.

\bibitem{rippel2021elf}
O.~Rippel, A.~G. Anderson, K.~Tatwawadi, S.~Nair, C.~Lytle, and L.~Bourdev,
  ``Elf-vc: Efficient learned flexible-rate video coding,'' \emph{arXiv
  preprint arXiv:2104.14335}, 2021.

\bibitem{wu2018video}
C.-Y. Wu, N.~Singhal, and P.~Krahenbuhl, ``Video compression through image
  interpolation,'' in \emph{Proceedings of the European Conference on Computer
  Vision (ECCV)}, 2018, pp. 416--431.

\bibitem{djelouah2019neural}
A.~Djelouah, J.~Campos, S.~Schaub-Meyer, and C.~Schroers, ``Neural inter-frame
  compression for video coding,'' in \emph{Proceedings of the IEEE/CVF
  International Conference on Computer Vision}, 2019, pp. 6421--6429.

\bibitem{yang2020learning}
R.~Yang, F.~Mentzer, L.~V. Gool, and R.~Timofte, ``Learning for video
  compression with hierarchical quality and recurrent enhancement,'' in
  \emph{Proceedings of the IEEE/CVF Conference on Computer Vision and Pattern
  Recognition}, 2020, pp. 6628--6637.

\bibitem{Pourreza_2021_ICCV}
R.~Pourreza and T.~Cohen, ``Extending neural p-frame codecs for b-frame
  coding,'' in \emph{Proceedings of the IEEE/CVF International Conference on
  Computer Vision (ICCV)}, October 2021, pp. 6680--6689.

\bibitem{habibi1974hybrid}
A.~Habibi, ``Hybrid coding of pictorial data,'' \emph{IEEE Transactions on
  Communications}, vol.~22, no.~5, pp. 614--624, 1974.

\bibitem{lindeberg2013scale}
T.~Lindeberg, \emph{Scale-space theory in computer vision}.\hskip 1em plus
  0.5em minus 0.4em\relax Springer Science \& Business Media, 2013, vol. 256.

\bibitem{lin2017feature}
T.-Y. Lin, P.~Doll{\'a}r, R.~Girshick, K.~He, B.~Hariharan, and S.~Belongie,
  ``Feature pyramid networks for object detection,'' in \emph{Proceedings of
  the IEEE conference on computer vision and pattern recognition}, 2017, pp.
  2117--2125.

\bibitem{zhu2020deformable}
X.~Zhu, W.~Su, L.~Lu, B.~Li, X.~Wang, and J.~Dai, ``Deformable detr: Deformable
  transformers for end-to-end object detection,'' in \emph{International
  Conference on Learning Representations}, 2020.

\bibitem{Liu_2021_ICCV}
Z.~Liu, Y.~Lin, Y.~Cao, H.~Hu, Y.~Wei, Z.~Zhang, S.~Lin, and B.~Guo, ``Swin
  transformer: Hierarchical vision transformer using shifted windows,'' in
  \emph{Proceedings of the IEEE/CVF International Conference on Computer Vision
  (ICCV)}, October 2021, pp. 10\,012--10\,022.

\bibitem{ladune2021conditional}
T.~Ladune, P.~Philippe, W.~Hamidouche, L.~Zhang, and O.~D{\'e}forges,
  ``Conditional coding for flexible learned video compression,'' in
  \emph{International Conference on Learning Representations (ICLR) 2021,
  Neural Compression Workshop}, 2021.

\bibitem{balle2016end}
J.~Ball{\'{e}}, V.~Laparra, and E.~P. Simoncelli, ``End-to-end optimized image
  compression,'' in \emph{5th International Conference on Learning
  Representations, {ICLR} 2017}, 2017.

\bibitem{guo2021soft}
Z.~Guo, Z.~Zhang, R.~Feng, and Z.~Chen, ``Soft then hard: Rethinking the
  quantization in neural image compression,'' in \emph{Proceedings of the 38th
  International Conference on Machine Learning}, vol. 139.\hskip 1em plus 0.5em
  minus 0.4em\relax PMLR, 2021, pp. 3920--3929.

\bibitem{mercat2020uvg}
A.~Mercat, M.~Viitanen, and J.~Vanne, ``Uvg dataset: 50/120fps 4k sequences for
  video codec analysis and development,'' in \emph{Proceedings of the 11th ACM
  Multimedia Systems Conference}, 2020, pp. 297--302.

\bibitem{wang2004image}
Z.~Wang, A.~C. Bovik, H.~R. Sheikh, and E.~P. Simoncelli, ``Image quality
  assessment: from error visibility to structural similarity,'' \emph{IEEE
  transactions on image processing}, vol.~13, no.~4, pp. 600--612, 2004.

\bibitem{wang2016mcl}
H.~Wang, W.~Gan, S.~Hu, J.~Y. Lin, L.~Jin, L.~Song, P.~Wang, I.~Katsavounidis,
  A.~Aaron, and C.-C.~J. Kuo, ``Mcl-jcv: a jnd-based h. 264/avc video quality
  assessment dataset,'' in \emph{2016 IEEE International Conference on Image
  Processing (ICIP)}.\hskip 1em plus 0.5em minus 0.4em\relax IEEE, 2016, pp.
  1509--1513.

\bibitem{minnen2018joint}
D.~Minnen, J.~Ball{\'{e}}, and G.~Toderici, ``Joint autoregressive and
  hierarchical priors for learned image compression,'' in \emph{Advances in
  Neural Information Processing Systems 31}, 2018, pp. 10\,771--10\,780.

\bibitem{Lee2019Context}
J.~Lee, S.~Cho, and S.~Beack, ``Context-adaptive entropy model for end-to-end
  optimized image compression,'' in \emph{7th International Conference on
  Learning Representations, {ICLR} 2019}, 2019.

\bibitem{chen2021end}
T.~Chen, H.~Liu, Z.~Ma, Q.~Shen, X.~Cao, and Y.~Wang, ``End-to-end learnt image
  compression via non-local attention optimization and improved context
  modeling,'' \emph{IEEE Transactions on Image Processing}, vol.~30, pp.
  3179--3191, 2021.

\bibitem{agustsson2017soft}
E.~Agustsson, F.~Mentzer, M.~Tschannen, L.~Cavigelli, R.~Timofte, L.~Benini,
  and L.~V. Gool, ``Soft-to-hard vector quantization for end-to-end learning
  compressible representations,'' in \emph{Advances in Neural Information
  Processing Systems 30}, 2017, pp. 1141--1151.

\bibitem{yang2020improving}
Y.~Yang, R.~Bamler, and S.~Mandt, ``Improving inference for neural image
  compression,'' in \emph{Advances in Neural Information Processing Systems},
  vol.~33, 2020, pp. 573--584.

\bibitem{agustsson2020universally}
E.~Agustsson and L.~Theis, ``Universally quantized neural compression,'' in
  \emph{Advances in Neural Information Processing Systems}, vol.~33, 2020, pp.
  12\,367--12\,376.

\bibitem{choi2019variable}
Y.~Choi, M.~El-Khamy, and J.~Lee, ``Variable rate deep image compression with a
  conditional autoencoder,'' in \emph{Proceedings of the IEEE International
  Conference on Computer Vision}, 2019, pp. 3146--3154.

\bibitem{cui2021asymmetric}
Z.~Cui, J.~Wang, S.~Gao, T.~Guo, Y.~Feng, and B.~Bai, ``Asymmetric gained deep
  image compression with continuous rate adaptation,'' in \emph{Proceedings of
  the IEEE/CVF Conference on Computer Vision and Pattern Recognition}, 2021,
  pp. 10\,532--10\,541.

\bibitem{agustsson2019generative}
E.~Agustsson, M.~Tschannen, F.~Mentzer, R.~Timofte, and L.~V. Gool,
  ``Generative adversarial networks for extreme learned image compression,'' in
  \emph{Proceedings of the IEEE/CVF International Conference on Computer
  Vision}, 2019, pp. 221--231.

\bibitem{yilmaz2021end}
M.~A. Y{\i}lmaz and A.~M. Tekalp, ``End-to-end rate-distortion optimized
  learned hierarchical bi-directional video compression,'' \emph{IEEE
  Transactions on Image Processing}, vol.~31, pp. 974--983, 2021.

\bibitem{feng2021versatile}
R.~Feng, Z.~Guo, Z.~Zhang, and Z.~Chen, ``Versatile learned video
  compression,'' \emph{arXiv preprint arXiv:2111.03386}, 2021.

\bibitem{li2021deep}
J.~Li, B.~Li, and Y.~Lu, ``Deep contextual video compression,'' \emph{arXiv
  preprint arXiv:2109.15047}, 2021.

\bibitem{lu2020content}
G.~Lu, C.~Cai, X.~Zhang, L.~Chen, W.~Ouyang, D.~Xu, and Z.~Gao, ``Content
  adaptive and error propagation aware deep video compression,'' in
  \emph{European Conference on Computer Vision}.\hskip 1em plus 0.5em minus
  0.4em\relax Springer, 2020, pp. 456--472.

\bibitem{sheng2022temporal}
X.~Sheng, J.~Li, B.~Li, L.~Li, D.~Liu, and Y.~Lu, ``Temporal context mining for
  learned video compression,'' \emph{IEEE Transactions on Multimedia}, 2022.

\bibitem{hu2022coarse}
Z.~Hu, G.~Lu, J.~Guo, S.~Liu, W.~Jiang, and D.~Xu, ``Coarse-to-fine deep video
  coding with hyperprior-guided mode prediction,'' in \emph{Proceedings of the
  IEEE/CVF Conference on Computer Vision and Pattern Recognition}, 2022, pp.
  5921--5930.

\bibitem{li2022hybrid}
J.~Li, B.~Li, and Y.~Lu, ``Hybrid spatial-temporal entropy modelling for neural
  video compression,'' in \emph{Proceedings of the 30th ACM International
  Conference on Multimedia}, 2022, pp. 1503--1511.

\bibitem{feng2020learned}
R.~Feng, Y.~Wu, Z.~Guo, Z.~Zhang, and Z.~Chen, ``Learned video compression with
  feature-level residuals,'' in \emph{Proceedings of the IEEE/CVF Conference on
  Computer Vision and Pattern Recognition Workshops}, 2020, pp. 120--121.

\bibitem{dai2017deformable}
J.~Dai, H.~Qi, Y.~Xiong, Y.~Li, G.~Zhang, H.~Hu, and Y.~Wei, ``Deformable
  convolutional networks,'' in \emph{Proceedings of the IEEE international
  conference on computer vision}, 2017, pp. 764--773.

\bibitem{zhu2019deformable}
X.~Zhu, H.~Hu, S.~Lin, and J.~Dai, ``Deformable convnets v2: More deformable,
  better results,'' in \emph{Proceedings of the IEEE/CVF Conference on Computer
  Vision and Pattern Recognition}, 2019, pp. 9308--9316.

\bibitem{bengio2013estimating}
Y.~Bengio, N.~L{\'e}onard, and A.~Courville, ``Estimating or propagating
  gradients through stochastic neurons for conditional computation,''
  \emph{arXiv preprint arXiv:1308.3432}, 2013.

\bibitem{mentzer2021towards}
F.~Mentzer, E.~Agustsson, J.~Ball{\'e}, D.~Minnen, N.~Johnston, and
  G.~Toderici, ``Towards generative video compression,'' \emph{arXiv preprint
  arXiv:2107.12038}, 2021.

\bibitem{xue2019video}
T.~Xue, B.~Chen, J.~Wu, D.~Wei, and W.~T. Freeman, ``Video enhancement with
  task-oriented flow,'' \emph{International Journal of Computer Vision}, vol.
  127, no.~8, pp. 1106--1125, 2019.

\bibitem{kingma2014adam}
D.~P. Kingma and J.~Ba, ``Adam: A method for stochastic optimization,''
  \emph{arXiv preprint arXiv:1412.6980}, 2014.

\bibitem{VTM}
VTM, ``Vvc offical test model,'' \url{https://jvet.hhi.fraunhofer.de}, 2021.

\bibitem{HM}
HM, ``Hevc offical test model,'' \url{https://hevc.hhi.fraunhofer.de}, 2021.

\bibitem{bjontegaard2001calculation}
G.~Bjontegaard, ``Calculation of average psnr differences between rd-curves,''
  \emph{VCEG-M33}, 2001.

\bibitem{lu2020end}
G.~Lu, X.~Zhang, W.~Ouyang, L.~Chen, Z.~Gao, and D.~Xu, ``An end-to-end
  learning framework for video compression,'' \emph{IEEE transactions on
  pattern analysis and machine intelligence}, vol.~43, no.~10, pp. 3292--3308,
  2020.

\end{thebibliography}

\end{document}